\RequirePackage{fix-cm}
\documentclass[twocolumn]{svjour3}  
\smartqed  
\usepackage{graphicx}
\usepackage{cite}
\usepackage[utf8]{inputenc}
\usepackage{amsmath,amssymb,amsfonts}
\usepackage{algorithmic}
\usepackage{enumerate}
\usepackage{glossaries}
\usepackage{graphicx}
\usepackage{mwe}
\usepackage{textcomp}
\usepackage{xcolor}
\usepackage{verbatim} 
\usepackage{paralist}
\usepackage{hyperref}
\usepackage{tabularx}
\usepackage{booktabs}
\usepackage{tabulary}

\newacronym{ACT-R}{ACT-R}{Adaptive Control of Thought-Rational}
\newacronym{AI}{AI}{Artificial Intelligence}
\newacronym{AutoML}{AutoML}{Automated Machine Learning}
\newacronym{BDP}{BDP}{Big Data Platform}
\newacronym{CBDP}{CAAI-BDP}{CAAI-Big Data Platform} 
\newacronym{CL}{CL}{Conceptual Layer}
\newacronym{CPPS}{CPPS}{Cyber-physical Production Systems}
\newacronym{DPL}{DPL}{Data Processing Layer}
\newacronym{CAAI}{CAAI}{Cognitive Architecture for Artificial Intelligence in Cyber-physical Production Systems}
\newacronym{HMI}{HMI}{Human-Machine Interaction}
\newacronym{I4.0}{I4.0}{Industry 4.0}
\newacronym{IoT}{IoT}{Internet of Things}
\newacronym{IIRA}{IIRA}{Industrial Internet Reference Architecture}
\newacronym{LOOCV}{LOOCV}{Leave-One-Out Cross Validation}
\newacronym{OPCUA}{OPC UA}{Open Platform Communications Unified Architecture}
\newacronym{RAMI4.0}{RAMI4.0}{Reference Architecture Model Industrie 4.0\-}
\newacronym{RMSE}{RMSE}{Root Mean Squared Error}
\newacronym{SMBO}{SMBO}{Surrogate Model-Based Optimization}
\newacronym{SME}{SME}{Small and Medium-sized Enterprise}
\newacronym{VPS}{VPS}{Versatile Production System}

\graphicspath{{figures.d/}}

\def\makeheadbox{\relax}

\journalname{Name of the Journal will be inserted after the review process completed}

\begin{document}

\title{CAAI---A Cognitive Architecture to Introduce Artificial Intelligence in Cyber-Physical Production Systems
}

\titlerunning{Cognitive Architecture to Introduce AI in Cyber-Physical Production Systems}        

\author{Andreas Fischbach  	\and
            Jan Strohschein		\and 
            Andreas Bunte		\and 
            J\"org Stork 			\and
            Heide Faeskorn-Woyke	\and 
            Natalia Moriz 		\and 
            Thomas Bartz-Beielstein}


\institute{  A. Fischbach 	\and 
		J. Stork 		\and
		T. Bartz-Beielstein \at
		TH Köln, Institute for Data Science, Engineering, and Analytics, Gummersbach, Germany, 
		\email{andreas.fischbach@th-koeln.de, joerg.stork@th-koeln.de, thomas.bartz-beielstein@th-koeln.de}
           	\and
		J. Strohschein \and
		H. Faeskorn-Woyke \at
		TH Köln, Institute of Computer Science, Gummersbach, Germany, 
		\email{jan.strohschein@th-koeln.de, heide.faeskorn-woyke@th-koeln.de}		
		\and
		A. Bunte \and 
		N. Moriz \at
		OWL University of Applied Sciences and Arts, Institute Industrial IT, Lemgo, Germany,
		\email{andreas.bunte@th-owl.de, natalia.moriz@th-owl.de}
}

\date{\vspace{3em}}

\maketitle
\begin{abstract}
This paper introduces CAAI, a novel cognitive architecture for artificial intelligence in cyber-physical production systems.
The goal of the architecture is to reduce the implementation effort for the usage of artificial intelligence algorithms. 
The core of the CAAI is a cognitive module that processes declarative goals of the user, selects suitable models and algorithms, and creates a configuration for the execution of a processing pipeline on a big data platform. 
Constant observation and evaluation against performance criteria assess the performance of pipelines for many and varying use cases.
Based on these evaluations, the pipelines are automatically adapted if necessary.
The modular design with well-defined interfaces enables the reusability and extensibility of pipeline components.
A big data platform implements this modular design supported by technologies such as Docker, Kubernetes, and Kafka for virtualization and orchestration of the individual components and their communication. 
The implementation of the architecture is evaluated using a real-world use case.

\end{abstract}

\section{Introduction}\label{sec:introduction}
The use of \gls{AI} in \gls{CPPS} can help to significantly reduce costs and provides new market opportunities~\cite{Kagermann:2013}. 
Many \gls{I4.0} applications rely on the use of \gls{AI}, such as condition monitoring, predictive maintenance, diagnosis, or optimization~\cite{FOF:2013,VBS:2017}.
Up until now, the implementation of real-world use cases is time and cost intensive due to missing standards and imprecisely described architectures.
Typically, an \gls{AI} expert analyzes one specific application and develops a suitable solution that will match customer needs. 
Often use cases, particularly those in \glspl{SME}, are not implemented because of limited resources and the unpredictable benefit of \gls{AI} solutions.
Therefore, to enable a resource-efficient use for many applications, \gls{AI} solutions require moderate manual implementation and operation effort.

Our goals (G) for the application of the \gls{CAAI} are represented by several requirements from the \gls{SME}s that have to be fulfilled by an \gls{AI} solution.

\begin{enumerate}[(G-1)]
	\item \textbf{Reliability:} 
	In a competitive market environment, the efficiency of \gls{CPPS} is important, and reliability is a prerequisite to achieving it.
	Since the \gls{CAAI} supports the \gls{CPPS}, both are interconnected and share the same requirements.  
	Capturing the complete data is essential, as it contains a vast amount of value, especially if it includes information about the quality of produced products, which can be utilized by applications of \gls{AI}. 
	To avoid losses of data or downtime of the \gls{CPPS}, the \gls{CAAI} and its underlying infrastructure have to be stable and reliable~\cite{Schroeder:2016, Drath:2014}.
	\item \textbf{Flexibility:} A significant drawback of existing \gls{AI} solutions is that an \gls{AI} expert often develops them for a single machine or a single problem. 
	Therefore these solutions do not include common interfaces, which enable an adaption or extension of the existing system. 
	This inflexibility is not acceptable, because there is a great demand in the market for fast adapting \gls{CPPS}, which can not be fulfilled by the approach of specific \gls{AI} solutions. 
	The \gls{CAAI} has to be flexible and extendable to enable a quick reaction on this market demand~\cite{Bunte:2019a}.
	\item \textbf{Generalizability:} The \gls{CAAI} should apply to many types of \gls{CPPS} and support many different use cases.
	As it is not possible to choose algorithms for all system type combinations and applications in advance, their selection should be performed automatically in an intelligent manner. 
	Thus the \gls{CAAI} has to process the user defined aims, derive a valuable process pipeline for the specific system, and learn over time to improve the system's performance, i.e., the \gls{CAAI} implements \textit{cognitive} capabilities~\cite{Bmwi:2019}.
	
	\item \textbf{Adaptability:} 
	The realization of adaptability through the \gls{CAAI} increases the efficiency of the \gls{CPPS} by directly adjusting process parameters, so that users do not have to change them manually. 
	Furthermore, adaptability allows to automate the adjustment, which is less error prone, and ultimately realize an autonomous system.
	However, the \gls{CAAI} has to ensure the safe operation of the \gls{CPPS} during the whole process. 
	For example, the operation boundaries of the \gls{CPPS} have to be respected during optimization, whereas in anomaly detection, there might be machine parts that need to proceed operation, even in a case of emergency. 
	Thus, the boundary conditions must be included in the \gls{CAAI} and \gls{CPPS} adjustments are only allowed within these boundaries~\cite{Bmwi:2019, Li:2019}.

\end{enumerate}
First, we review existing approaches. They are associated with the goals \mbox{(G1-G4)}, but up until now, no work tackles those goals properly.
Architectures, which were introduced in the domain of automation, such as the \gls{RAMI4.0}~\cite{VDI:2015}, the \gls{IIRA}~\cite{Lin2017}, or the 5C architecture~\cite{Lee:2017}, are too abstract since they do not define implementation details, such as interfaces.
To achieve a more specific architecture, it is necessary to extend and refine certain parts of them.
Cognitive architectures, such as \gls{ACT-R}~\cite{Anderson:1996} and Soar~\cite{Laird:1987}, implement certain concepts to reach adaption and cognitive capabilities.

They can not be directly used to address industrial use cases, because they focus on cognition and lack of generality~\cite{BFS:2019}.
\gls{AutoML}~\cite{Feur15a} and hyperheuristics can choose and configure a suitable algorithm automatically. 
That includes steps such as data pre-processing, algorithm selection, and hyperparameter optimization~\cite{Thor13a, Fusi18a, Olso16a}.
Since there is an intersection between our architecture and these methods, they are considered for our implementation.
Some \glspl{BDP} can be found in the literature, such as the Open Big Data Platform for Industry 4.0~\cite{Weskamp:2019}, the Big Data Analytics Architecture for Industry 4.0~\cite{Santos2017}, and the Big Data for Industry 4.0~\cite{Gokalp2016}.
To the best of our knowledge, none of these \glspl{BDP} fulfills all our requirements towards a cognitive architecture. 

According to Neisser~\cite{Neis14a}, cognition refers to “all processes by which the sensory input is transformed, reduced, elaborated, stored, recovered, and used”.
Regarding the context of \gls{I4.0}, in the scope of \gls{CAAI} we define cognition as follows.
\begin{definition}[Cognition]\label{def:cognition}
\textit{Cognition} refers to all processes by which the input data is transformed, reduced, elaborated, stored, recovered, and used to solve \gls{I4.0} use cases, i.e., condition monitoring, anomaly detection, optimization, and predictive maintenance.
\end{definition}
The central part of \gls{CAAI} is a cognitive module, which processes knowledge, interprets aims, creates appropriate pipelines, and improves the system by continuous evaluation.
Existing technologies from \gls{AutoML} and hyperheuristics can be integrated into the cognitive module to create pipelines.
Additionally, the cognition module stores \emph{a priori}\/ knowledge that is valid for all use cases, e.g., information about suitable algorithms to solve specific tasks such as multi-criteria optimization or time series anomaly detection. 
By learning from experience, the knowledge grows and the performance of the system improves over time.
Due to its modularity, the architecture is extensible, and allows the integration of new algorithms into the \gls{CAAI}. 
Furthermore, the architecture enables an adaption of the \gls{CPPS} if a promising configuration was determined.
Even though it is difficult for \gls{CAAI} to reach results that are equivalent to a customized solution from an \gls{AI} expert, it will achieve improvements in a cost-efficient manner for many use cases, without the need of support from an \gls{AI} expert.

Depending on the individual problem characteristic, which changes from use case to use case, some algorithms might be superior to others in terms of either performance or computation time. 
Additionally, each algorithm needs several pre-processing steps, e.g., feature creation, feature selection, or model building. 
The type of models, their parameter values and the algorithm topologies can be learned, at least to some extent, by the system itself. 

The contribution of this paper is a novel cognitive architecture for \gls{CPPS}, which has several advantages in comparison to the state-of-the-art architectures. 
To tackle goals \mbox{(G-1)} to \mbox{(G-4)}, the  following methods (M) are considered in this paper:
\begin{enumerate}[(M-1)]
	\item \textbf{Big Data Platform:} 
	Continuous and reliable operation \mbox{(G-1)} is ensured by a \gls{BDP}.
	The \gls{BDP} comprises different techniques to reach the goal, such as orchestration, virtualization, and containerization. 
	The orchestration instantiates and connects the selected modules and thus creates the pipelines needed for the processing.
	Furthermore, the orchestration enables to move applications and their respective containers to the remaining infrastructure if certain parts of the system fail.
	Containerization, which provides virtualization on operating-system-level, allows the existence of multiple isolated user-space instances.
	It improves reliability because each instance can only access its container's contents and devices assigned to this container~\cite{Negu15a}. 
	Orchestration of virtualized containers also aids scalability as it is possible to create several instances of a container to work in parallel.
	
	\item \textbf{Modularization:} 
	A modularization of the \gls{AI} components enables the  flexibility \mbox{(G-2)} of the \gls{CAAI} to adapt it to specific requirements. 
	Moreover, modular components require well-specified interfaces with detailed definitions.
	Furthermore, the modular design reduces development and maintenance costs by the integration of existing components. 
	Thus, only new components have to be developed.

\item \textbf{Cognition:} Automated process pipeline generation methods enable the realization of different use cases without the involvement of an \gls{AI} expert and the spreading to different types of \gls{CPPS}.
The automated pipeline creation is an important feature to ensure generalizability \mbox{(G-3)} for specific use cases, dynamic systems, and changing environments.
It is realized by the cognitive module, which generates the pipeline and selects the best fitting algorithms according to the given data and defined goal. 
The pipeline evaluation is automated to improve results and detect performance drifts. 
The cognitive module subsequently evaluates the various pipelines to collect information about the performance of different algorithms for a given use case.
This evaluation enables learning over time and the detection of performance drifts that may result in a re-calibration of the processing pipeline.

\item \textbf{Automatic Decision:} 
Knowledge is needed to interpret the algorithms results and derive suggestions to realize automatic or \gls{CAAI} supported adaptions.
Therefore, new parameter sets or other system changes, identified by the algorithms, have to be applied to the \gls{CPPS} \mbox{(G-4)}. 
Furthermore, boundary conditions can be defined and applied for decision making, e.g., a minimum of expected improvement or specifications that ensure a safe adaption.
Moreover, the decision must be applied to the \gls{CPPS} controller, which adapts the physical machine.
For realization, several existing approaches, such as skill-based engineering, can be used \cite{Bunte:2019a, Zimmermann:2019, Malakuti:2018}.
\end{enumerate}

The remainder of this paper is organized as follows:
Section~\ref{sec:architecture} introduces the proposed cognitive architecture. 
A real-world use case evaluates the \gls{CAAI} architecture as described in Section~\ref{sec:implementation} along with available techniques to implement the methods, design choices, and the results of our implementation. 
Finally, the conclusion and outlook Section~\ref{sec:conclusion} presents our major findings and resulting future research tasks.

\section{The \gls{CAAI} Architecture} \label{sec:architecture}
In this section, we introduce the \gls{CAAI}, which aims to have a reference character, on a rather abstract level.
Detaching the description from the concrete implementation allows choosing implementation technologies according to individual preferences or technological advances.
Nevertheless, Section \ref{sec:implementation} introduces an exemplary implementation of the \gls{CAAI}.
The architecture address\-es the methods \mbox{(M-1)} to \mbox{(M-4)} to reach the goals \mbox{(G-1)} to \mbox{(G-4)}.

\subsection{Overview}
\begin{figure*}
\centering
\includegraphics[width=0.65\textwidth]{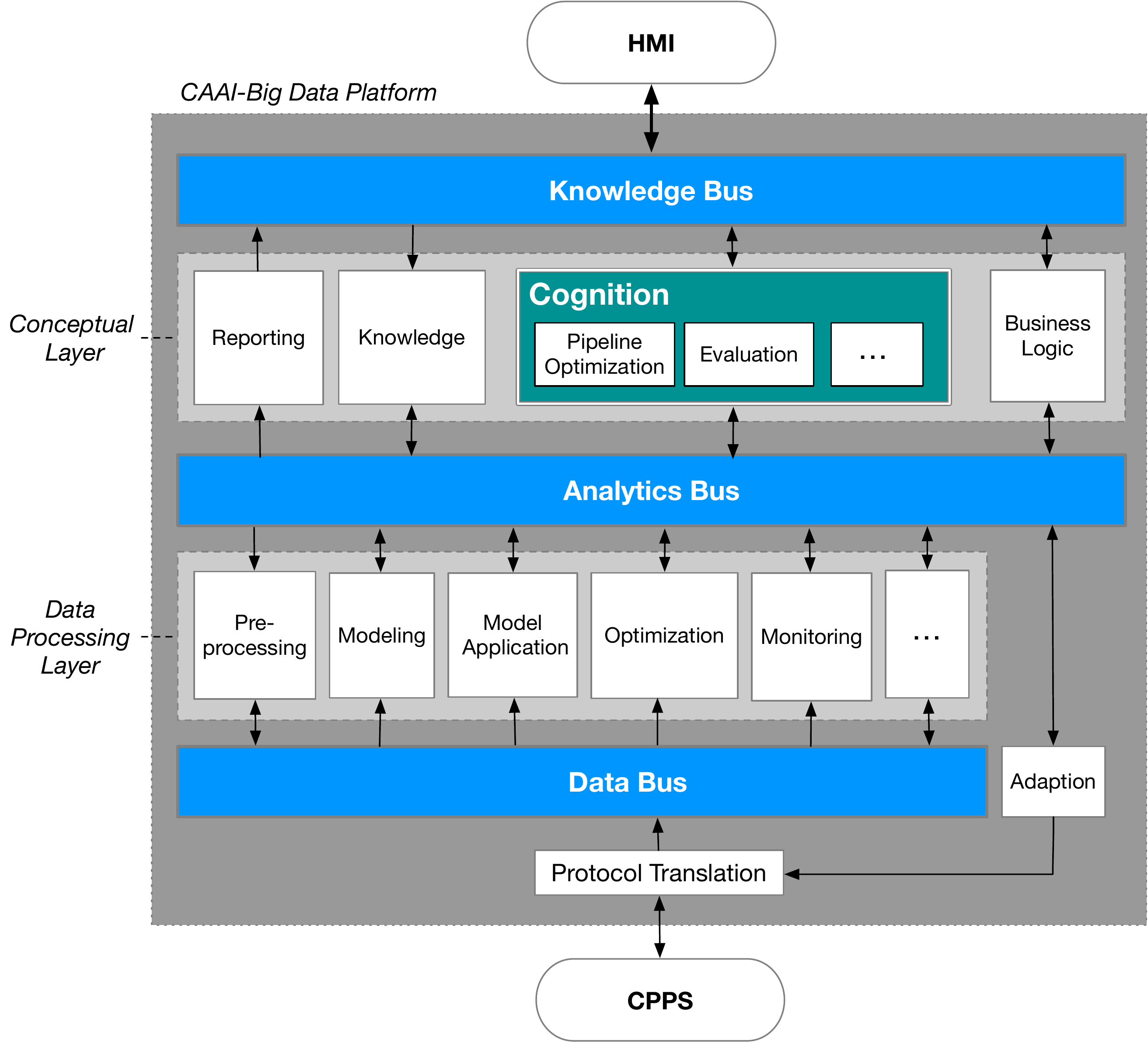}
\caption{\gls{CAAI} architecture overview. The \gls{CBDP}, depicted in dark grey, manages the bus systems and layers.
Bus systems are colored in \emph{blue} and establish communication between the modules. The arrows demonstrate the designated information flow. 
The layers are shaded in light grey and contain the different modules. 
The cognition module, which establishes automatic configuration, is colored in \emph{turquoise}. 
Oval shapes depict external systems, e.g., a \gls{CPPS} or a Human Machine Interface.}
\label{fig:architecture}
\end{figure*}

Our cognitive architecture builds upon the idea of modeling the information, data applications, and streams required for specific tasks in the \gls{I4.0} scenario while providing reliability, flexibility, generalizability, and adaptability.
The concept, depicted in Figure~\ref{fig:architecture}, is based on a three-tier architecture to simplify interoperability and to ensure horizontal scalability. 

The \gls{CAAI} approach addresses sub-symbolic \gls{AI}. In contrast to symbolic \gls{AI}, sub-symbolic \gls{AI} does not use human-readable problem representations. 
Sub-symbolic \gls{AI}, such as neural networks and deep learning, perform calculations based on principles that have demonstrated to be able to solve problems.

The \gls{CBDP} wraps the architecture and arranges software modules in two processing layers, the \gls{DPL}, and the \gls{CL} and connects them via three bus systems (data, analytics, and knowledge bus). 
Data from a \gls{CPPS} enters the system at the very bottom (see Figure~\ref{fig:architecture}). 
The protocol translation module transforms incoming data and sends it to the data bus. 
The pre-processing module receives the raw data, performs the necessary steps to clean the data, and publishes the results back to the data bus. 
Other modules in the data processing layer utilize data from the data bus and transfer their analytical results onto the analytics bus. 
Modules in the \gls{CL} process information about the user defined aims and the business logic for a given use case. 
They evaluate the results from the analytics bus, determine the parameters to adjust the \gls{CPPS} via the adaption module, and measure the overall system performance. 
The \gls{CL} modules also interact with the knowledge bus to generate reports for the user and to process new instructions. 

The central element of the architecture is the cognitive module, which selects and orchestrates different analyses and algorithms, depending on the use case. 
Therefore the composition of active modules and their communication over the bus system will change during run time. 
Providing a pre-defined set of modules and the capability to add new modules reduces the overall implementation complexity by building a cohesive yet modular solution. 
The following sections describe the bus infrastructure, the layers, and the cognitive module in more detail.

\subsection{Bus Infrastructure}
Different modules of the system communicate asynchronously via three buses, the data bus, analytics bus, and knowledge bus. 
The processing degree of the data grows incrementally from bottom to top, and additionally in some cases horizontally as well.  
Each bus covers several topics, which can be subscribed by modules attached to the bus. 
All modules publish their data to the relevant attached bus on a pre-defined topic, so one or more other modules can use the data and intermediate results for their processing.  
Thus, the architecture implements a message-driven processing approach, leading to a flexible and agile system with clear interfaces and hierarchies. 
The main features of applications that use message queuing techniques are~\cite{ibmmq}:
\begin{enumerate}[(i)]
	\item no direct connections between modules,
	\item communication between modules can be independent of time,
	\item work can be carried out by small, self-contained modules,
	\item communication can be driven by events,
	\item data integrity through validation schemata, and
	\item recovery support.
\end{enumerate}
\begin{table*}\centering
    \caption{Comparison of the data on the three bus systems used in the \gls{CAAI}. 
    }
    \label{tab:bus}
\begin{tabular}{llll}
    \toprule 
    				 & Data bus & Analytics bus & Knowledge bus \\
    \midrule
    \midrule
	Type of data   & Raw, Pre-processed          & Processed       &      Enriched       \\
	\midrule
	Volume and Velocity & High & Moderate & Low \\
	\midrule
	Computational effort & Low		& High 	&  Moderate  \\
	\midrule
	Entropy  & Low & Moderate & High \\
	\midrule
	Interpretability & Complicated        &  Moderate     & Easy \\
    \bottomrule   
\end{tabular}
\end{table*}

The following paragraphs give a detailed description of the three bus systems, while Table~\ref{tab:bus} presents an overview and summarizes the differences of the respective data type.
\subsubsection*{Data Bus}
The data bus transports raw data from a \gls{CPPS}, as well as data from demonstrators, external simulators, or simulation modules.  
Cleaned and further pre-processed data is also published back to the data bus by the pre-processing modules.
Therefore the data volume and velocity is high, even though the entropy is still quite low, and interpretability is complicated. 
Overall the data bus transports streams of real-time data. 

\subsubsection*{Analytics Bus}
Data transported on the analytics bus have a higher information density than data from the data bus. 
The number of processing pipelines and the type of used algorithms determines the computational effort. 
These get instantiated by the cognition module and are expected to use a significant amount of the available processing power. 
Consequently, the analytics bus hosts knowledge, models, and results from the model application, the monitoring module, and the business logic to derive commands to adjust the system.

\subsubsection*{Knowledge Bus}
The knowledge bus enables the communication between the end-user and the system and combines the knowledge, business logic, and user-defined goals and actions.  
The cognitive module receives declarative goals defined by the user \cite{BFS:2019}.
Information from the analytics bus condenses into reports for the user. 
Furthermore, feedback from the user can be requested. 
So the knowledge bus uses enriched data, which aids the interpretability and provides the most value for the user.

\subsection{Layer}    
Several modules process the data within two layers. 
Each layer can be extended individually.
Each of the modules processes the data in a specific manner, e.g., with a specific algorithm.
Several modules are combined to enable complex data processing.

\subsubsection{Data Processing Layer}
\begin{figure}
	\centering
	\includegraphics[width=\columnwidth]{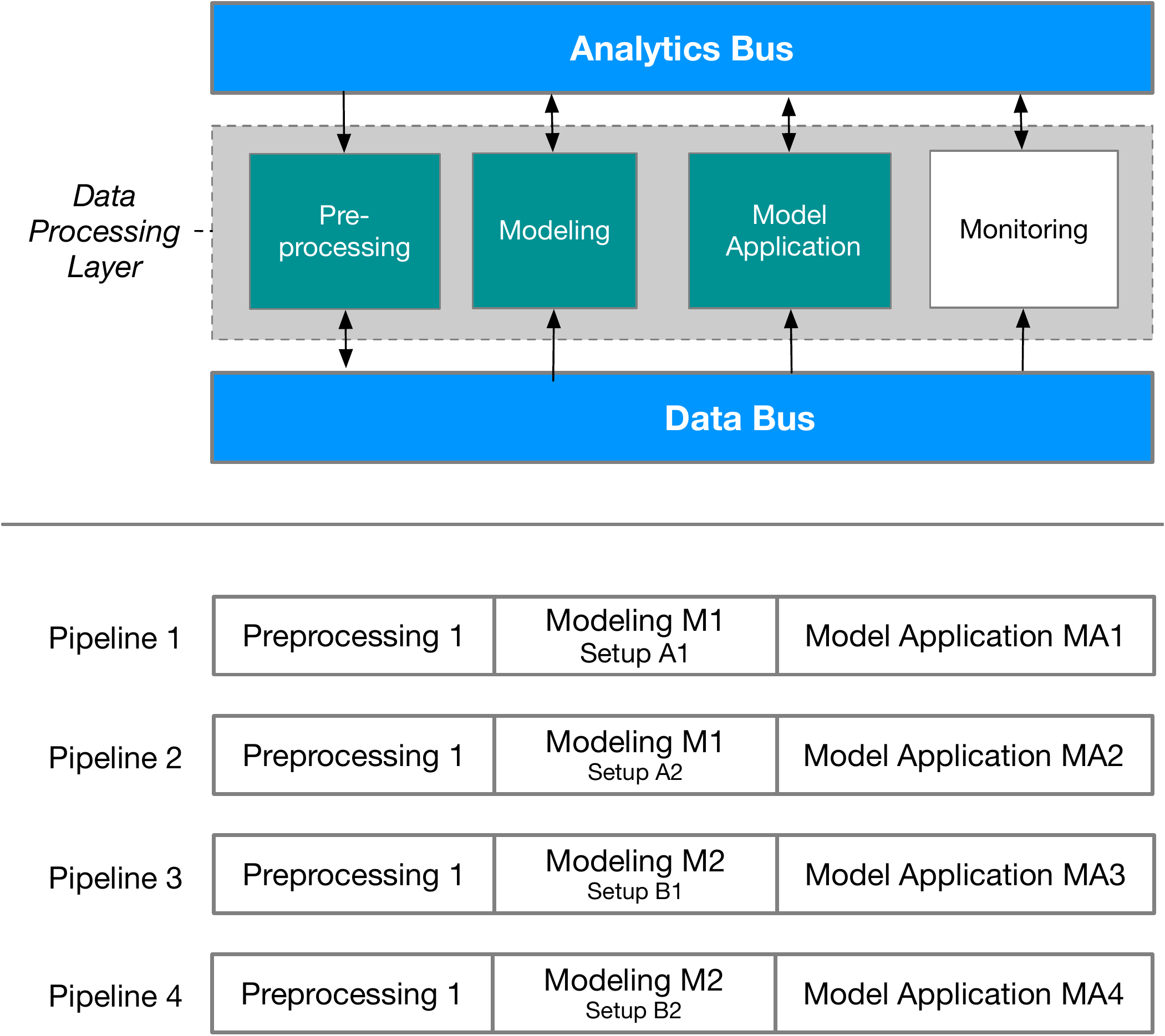}
	\caption{\emph{Top:}\/ \gls{DPL} with modules highlighted that can be selected and varied by the cognitive module.
	\emph{Bottom:}\/ An example of four different process pipeline candidates. Two different modelling Algorithms (M1, M2) with different parameter setups (Setup A1, A2, B1, B2) and four different Modeling Application modules (MA1 to MA4) are used. Here, all pipelines use shared results of one preprocessing module.}
	\label{fig:pipeline}
\end{figure}

The \gls{DPL} handles sub-symbolic data and therefore contains all modules that are processing sub-symbolic data.
Instances of modules are combined to processing pipe\-lines in order to solve a desired task based on the raw data (see Figure~\ref{fig:pipeline}).
Except of the monitoring module, all initially provided modules belong to one of the following three types:
\begin{enumerate}[(i)] 
	\item Pre-processing modules receive data from the \textit{data bus} and provide results to the \textit{data bus}. 
		They prepare data for the usage, e.g., by imputing missing values or synchronizing time stamps.
	\item Modeling modules receive pre-processed data from the \textit{data bus} and send their results to the \textit{analytics bus}.
	\item Model application modules get data from the \textit{data bus} and the \textit{analytics bus} and send their results to the \textit{analytics bus}.
\end{enumerate}
The modeling modules contain different machine learning algorithms in a modular manner.
There is a need that the architecture incorporates multiple algorithms to chose appropriate modules, based on the task, the type of data, and the resulting model.
It is also possible that modeling modules integrate some expert knowledge into the model and provides tools for data curation. 
Model application modules can access the final model on the \textit{analytics bus}.
Additionally, the model application will access the \textit{data bus} to compare the model with the process data to detect deviations, which are provided to the \textit{analytics bus}.
The cognitive module ensures that each model application module is compatible with a specific task and a particular model.
Each of these components has a particular purposes, such as condition monitoring, predictive maintenance, diagnosis, optimization, or similar tasks.

\subsubsection{Conceptual Layer}
The \gls{CL} is located between the \textit{analytics bus} and the \textit{knowledge bus} and contains the following four modules.
\begin{enumerate}[(i)]
	\item The reporting module visualizes the process data for the \gls{HMI}.
		It processes the data resulting from, e.g., monitoring or model application results. 
	\item The knowledge module contains 
		\begin{enumerate}[(a)]
			\item relevant information about the \gls{CPPS}, such as signal names, types of devices, or its topology, 
			\item general knowledge, such as an algorithms topology, which describes the ability and properties of algorithms, and
			\item constraints that can be defined by the user, e.g., time constraints.
		\end{enumerate}
	\item The business logic module decides whether an action is required or not.
	Therefore, it monitors the results from the model application modules, checks the constraints from the knowledge module, and derives actions, e.g., an adaption of the \gls{CPPS} when a certain threshold is reached. 
	\item The cognition module is responsible for the pipe\-line creation and optimization. 
		If a specific task is provided by the user, the cognitive module aggregate and configure suitable modules of the \gls{DPL} to fulfill the task.
		Hence it uses the algorithm topology of the knowledge module as well as past experiences.
		Monitoring the results of a specific aggregation enables the learning of functional aggregations and thus improves the performance over time.
		Therefore, the cognitive module is an elementary module of the \gls{CAAI} and the reason why it is called a cognitive architecture.
\end{enumerate}

\subsection{Cognition}
\begin{figure}
	\centering
	\includegraphics[width=1\columnwidth]{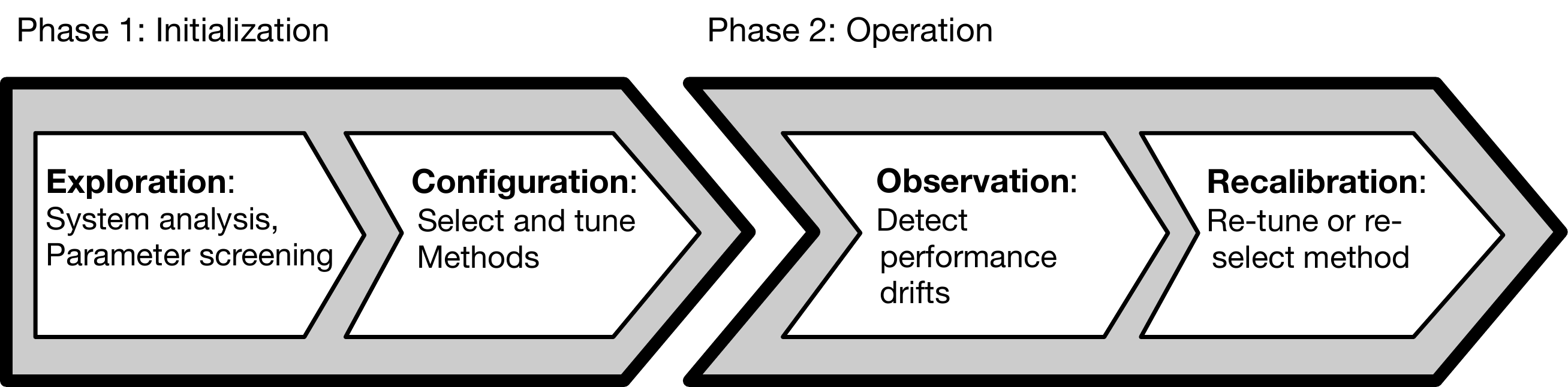}
	\caption{Two working phases of the cognition module: Phase 1 initializes the search for the best fitting pipeline utilizing a space-filling design like Latin Hypercube Sampling. During phase 2 the process is observed with respect to performance drifts. A re-calibration of the methods can be performed on demand.}
	\label{fig:cognition}
\end{figure}

The cognition module is a crucial part of the \gls{CAAI} architecture, as it enables the system to learn over time and transfer knowledge to several use cases (G-3).
It is responsible for major tasks in the \gls{CAAI} architecture, such as the algorithm selection, parameter tuning, and system management.
To properly address these tasks, the following preconditions have to be fulfilled:  
\begin{enumerate} [(i)]
	\item Feature engineering is the task of selecting and extracting relevant features from sensor data after or during the pre-processing.
	Involving domain knowledge and years of experience from the engineers is considered as a prerequisite and can significantly speed up process time and boost the quality of resulting models. 
	\item A declarative goal for the system has to be given, e.g., ''minimize energy consumption''. 
	Furthermore, the goal needs to be reflected in the \gls{CPPS} and the sensor data.
	A set of appropriate algorithms to address the specified goal has to be available.
	\item Finally, relevant knowledge and business logic to solve the given task must be available.
\end{enumerate}
As illustrated in Figure~\ref{fig:cognition}, the cognition module works in two phases, initialization and operation.

\subsubsection*{Phase-1: Initialization} 
The cognitive module chooses one or more processing pipelines.
Pipelines typically consist of pre-processing, modeling, and one or more model applications (such as classification, regression, or optimization).
While some modules may require special predecessors, e.g., a certain pre-processing, other module instances can be equal across different pipelines, and consequently, their results must be computed only once. 
The \gls{CBDP} orchestrates the sequence of modules and manages their processing, which might be in parallel.
The initialization phase utilizes automated selection tools such as \gls{AutoML} or hyperheuristics. 
We implement \gls{SMBO} to model the performance of algorithms and suggest new promising algorithm configurations by utilizing model-predictions~\cite{Bart16n,Thom18a}.
Once pipelines are selected, the initialization step includes the tuning of associated parameters of all included methods, to omit false configurations and wrong parameter settings. 
At the end of phase 1, the cognition evaluates the candidate pipelines and chooses the best according to their performance.
\subsubsection*{Phase-2: Operation} 
The cognition is responsible for the observation of the processing pipeline in an online manner to detect drifts or performance decreases during the operation phase. 
These can occur if circumstances change over time, e.g., the quality of a material used in the production process. 
When such situations appear, the cognition performs a re-calibration of the processing pipeline, which includes a new selection or reconfiguration of the modules. 
This feature allows the system to adapt to new situations in the production process automatically.
Moreover, the performance monitoring of the data structure itself and the performance of the chosen algorithms on the data enables the system to learn over time which methods are suitable to solve desired tasks. 

\section{Implementation}\label{sec:implementation}
In this section, we introduce the considered use case, followed by implementation details about the \gls{CBDP}.
Furthermore, we present the process description of how \gls{CAAI} behaves and introduce the results of our implementation.

\subsection{Use case}\label{sec:usecase}
We evaluate the \gls{CAAI} through its implementation for the \gls{VPS}, which is located in the SmartFactoryOWL.
The \gls{VPS} is a modular production system, which processes corn to produce popcorn which is used as packaging material.
Typically, there are four \gls{VPS} units, namely delivery, storage, dosing, and production. 
Due to its modularity, the first three units can be exchanged or removed easily. 
Depending on the current orders, different configurations are used.
The need for different configurations rises, e.g., if a small and exact amount of popcorn should be produced, which is performed by the dosing unit. 
However, if larger amounts are requested, it is more efficient to renounce the dosing unit because it is slow, and it generates operation costs.
Efficiently operating the \gls{VPS} is a challenge because many parameters influence the result, e.g., the moisture of the corn, the rate of corn that does not pop, or the amount of corn within the reactor.
Since not all parameters can be measured inline, data-driven optimization is a promising method to increase efficiency.
Therefore, the \gls{CAAI} architecture perfectly matches the requirements of the \gls{VPS} use case.
As a basis, a reliable, easy to set up, and scalable environment for the \gls{AI} is needed, which refers to method  \mbox{(M-1)}.
Since the configuration is regularly changing, the \gls{AI} components have to be modular \mbox{(M-2)},  re-useable, and extendable.

Due to reconfigurations of the \gls{VPS}, the use case and the \gls{VPS} units might change over time, hence there is the need for cognition \mbox{(M-3)}. 
However, improvements should be directly applied to the \gls{VPS} to reach the best performance, where automatic decisions \mbox{(M-4)} are needed.

In this use case, all \gls{VPS} units are used, and small boxes of popcorn are produced.
In each batch, one box of popcorn has to be filled. 
The overage of popcorn produced in one batch, or not fully filled boxes cannot be used, so it is waste.
Optimizing the amount of corn in the reactor, as provided by the dosing unit, is the goal.
The optimum is a trade-off between three minimization functions: the energy consumption ($f_1$), the processing time ($f_2$), and the amount of corn needed for a small box ($f_3$).
These functions are conflicting to some degree. 
The result of the optimization is a parameter value for the dosing unit.
The parameter $x$ controls the runtime of the conveyer and, therefore, indirectly influences the amount of corn processed.
As the given optimization problem can be regarded as relatively simple,
we will apply a single objective optimization algorithm and compute a weighted sum of the objectives.
This results in the following optimization problem:
\begin{equation}
\label{eq:prob}
\min ~\sum_{i=1}^{3} w_i f_i(x); \text{~~w.r.t~~}  w_i > 0 \text{ and } \sum_{i=1}^{3}w_i = 1
\end{equation}
The scalar weights of the corresponding objectives, $w_i$,  are chosen based on user's preferences. 
As a default, equal weights are used. 
The minimum of (\ref{eq:prob}) is a Pareto-optimal solution~\cite{Marler2010}.
The problem will be optimized by \gls{SMBO}~\cite{Jin05a}.
\gls{SMBO} utilizes a data-driven surrogate model to create an approximation of the real \gls{VPS} production process.
The model construction requires sampled data for a set of $n$ values of $x$, which should ideally depict a representative set of all possible settings, i.e., in a space-filling manner.
In this case, the set is generated by evaluating an equidistantly spaced design in the complete parameter range of $x$.
The cognition will evaluate different surrogate models: random forest~\cite{Brei01a} and Kriging~\cite{Krige1951}. 
Kriging is especially suitable for modeling continuous data with few variables and comes with an uncertainty measurement. 
At the same time, random forest is also able to model discrete parameters and computes very fast. 
Recent examples of applications of Kriging and random forest in \gls{CPPS} scenarios can be found in~\cite{Jung17a,Xing18a}.

\begin{figure}
	\centering
	\includegraphics[width=\columnwidth]{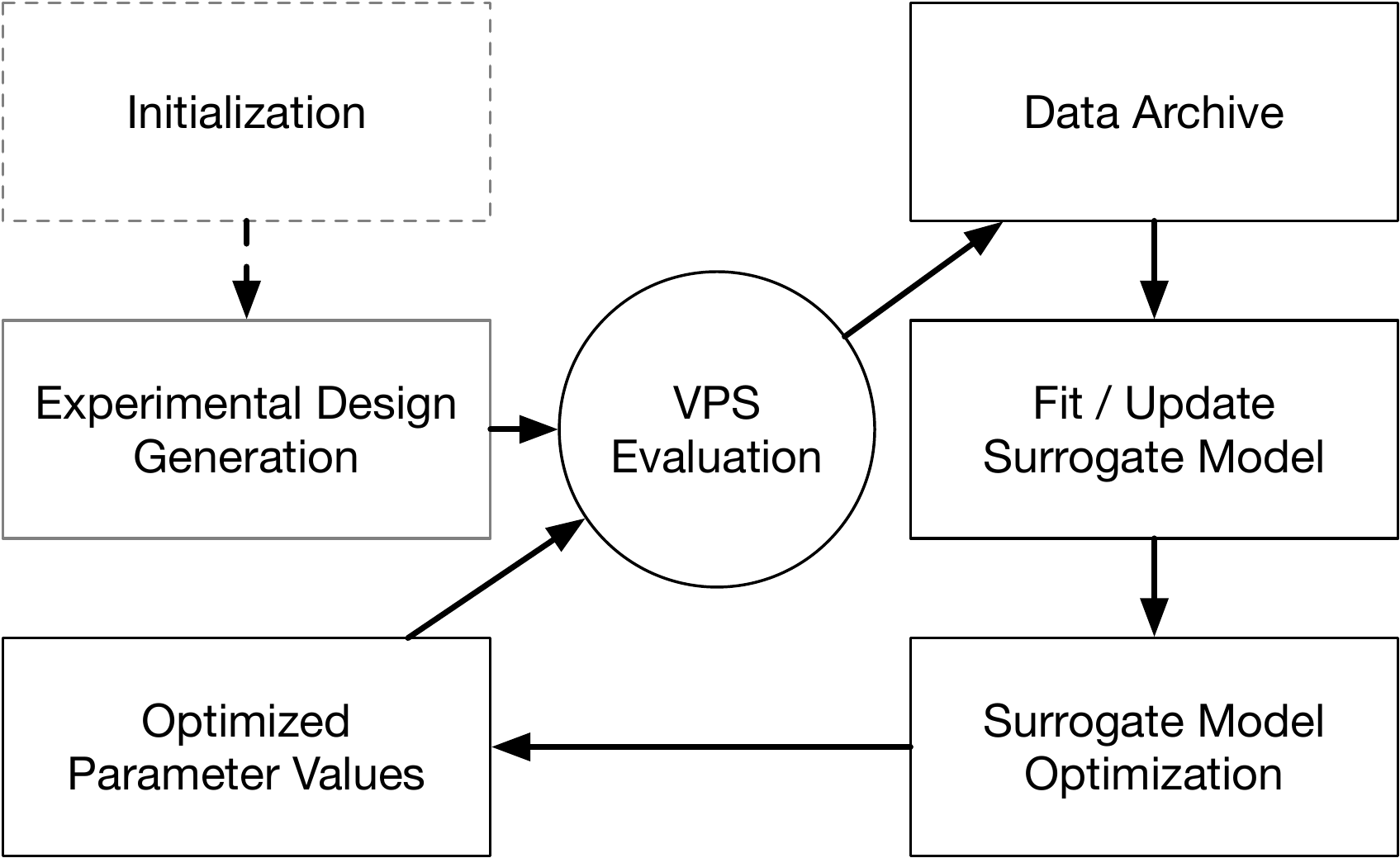}
	\caption{The SMBO optimization cycle starts in the upper left with the initialization and design generation. Then the loop is processed with the evaluation of the design, the model computation and optimization to retrieve the next candidate solution to be evaluated in the VPS.}
	\label{fig:smbo_cycle}
\end{figure}

With these two modeling algorithms, a broad range of systems can be covered. 
Furthermore, these surrogate models may differ in their hyperparameters, which results in a large number of possible configurations. 
The cognition decides which model and parameterization fit best to approximate the process data and perform optimization based on performance evaluation of the whole optimization cycle.
The surrogate will then be optimized to identify the next candidate solution to be evaluated on the \gls{VPS} by applying a local search algorithm.
Figure \ref{fig:smbo_cycle} shows the optimization cycle of \gls{SMBO}. 

If a new parameter has been identified the business logic defines when an adaption should be performed and verifies boundaries, such as limitations for parameters to protect the \gls{VPS}.
Finally, the adaption module adapts the \gls{VPS}, i.e., changes parameters to achieve better performance.

\subsection{CAAI - Big Data Platform}
\begin{figure}
	\centering
	\includegraphics[width=1\columnwidth]{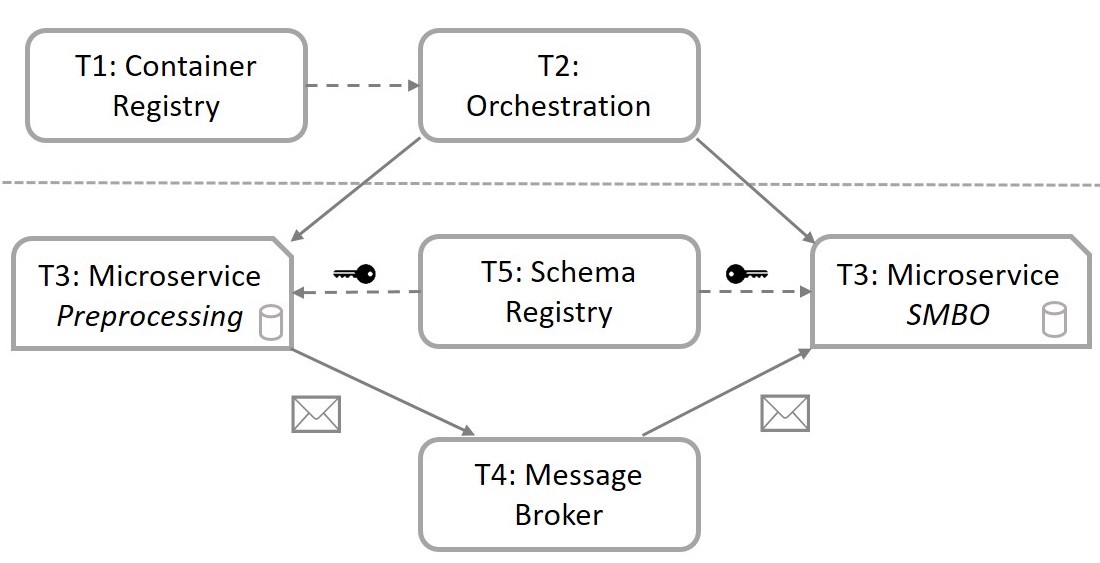}
	\caption{Cognitive Big Data Platform. Five technologies compose the \gls{CBDP}. Images of the different modules are stored in the container registry (T1). The orchestration (T2) uses these images to instantiate the message broker (T4) as well as the schema registry (T5), to enable standardized communication between modules. Following that, the cognition instructs the orchestration, which modules compose a data processing pipeline (T3).}
	\label{fig:big_data_platform}
\end{figure}

The \gls{CBDP} is a distributed system and can be hosted on a single machine, an on-premise cluster, or by a cloud provider.
Our stated architectural goals \mbox{(G1-G4)} are supported by the \gls{CBDP} through the implementation technologies, which are presented below.
Transferring incoming/outgoing data, orchestrating the data processing tasks, assuring the persistence of results, and managing the communication between modules are the resulting tasks of the \gls{CBDP}.
The following concepts and technologies (T) are used to accomplish the tasks. 
\begin{enumerate}[(T1)]
	\item \textbf{Container Virtualization:} 
	All components of the system exist as virtualized containers on the \gls{CBDP}. 
	Isolating the module requirements from the general environment ensures that all requirements for a specific module are met and do not interfere with other modules on the same platform, similar to virtual machines.
	In contrast to virtual machines, a container uses the host operating system, and containers share binaries and libraries if possible, which results in less overhead.
	Containers are consistent and immutable, which ensures compatibility across systems.
	A central container registry stores the container images and keeps track of changes via versioning.
	Docker is used as a container engine for the implementation of this use-case~\cite{Negu15a}.
	Images consist of all the necessary code instructions to install the requirements and create a specific environment in order to execute the desired algorithm or software.
	Generally speaking, a validated, running image guarantees to work the same on every computer, server, or cloud-environment. 
	\item \textbf{Orchestration:}
	The \gls{CBDP} manages the necessary infrastructure and orchestrates virtualized components to compose a system consisting of microservices that perform a specific task. 
	Orchestration frameworks handle deployments, configuration, updating, and removing of the virtualized software components. 
	A text file declaratively composes a system and lists the different services. 
	Orchestration is done by Kubernetes, which can utilize the Docker container engine~\cite{High17a}.
	The \textit{cognition} module uses the \textit{orchestration} to instantiate pipelines with selected algorithms and evaluate the results. 
	\item \textbf{Microservices:}
	All modules are developed as microservices to compose the software system for a specific use case from smaller self-sufficient parts. 
	Each module includes standardized communication functionality to publish and subscribe to relevant topics on the bus system~\cite{Watt15a}.
	The resulting system is modular, language-agnostic, and utilizes well-defined interfaces. 
	According to microservice best-practices, each microservice can store internal data in its local storage.
	\item \textbf{Messaging:}	The different bus systems managed by the \gls{CBDP} transfer data via messaging. 
	Messaging allows asynchronous communication between modules and enables parallelization as well as processing data several times for different purposes via topics and consumer groups. 
	Adding more instances to the same consumer group would result in a distributed processing of incoming messages, which is useful if a task is very time-consuming or response time is restricted.
	We chose Kafka~\cite{Nark17a} as a reliable message system for our platform.
	\item \textbf{Schema Management:}
	A schema stores the metadata of the data, with all the available fields and datatypes~\cite{Conf20a}. 
	When a module publishes to the bus system, the serializer applies the schema and encodes the message or filters out non-con\-for\-ming messages. 
	A consumer that subscribes to a topic on the bus has access to the same schema and can verify the integrity before encoding the incoming message. 
	Therefore clear communication via the bus system is ensured, and additional modules can be integrated easily. 
	A central schema registry distributes and versions the schemas which allow regulated data evolution.
\end{enumerate}
The combination of technologies \mbox{(T1-T5)} supports the overall goals and the methods to reach those, namely providing a reliable infrastructure \mbox{(M-1)} for modular development \mbox{(M-2)}, e.g., re-using existing modules or extending the system with additional algorithms. 
This enables the cognition to run and evaluate additional experiments through the automatic creation of processing pipelines \mbox{(M-3)} and an automatic adaption of the CPPS \mbox{(M-4)} if a feasible and beneficial solution was found. 
The interaction of the five technologies is illustrated in Figure~\ref{fig:big_data_platform}.

\subsection{Process Description}
\begin{figure}
	\centering
	\includegraphics[width=1\columnwidth]{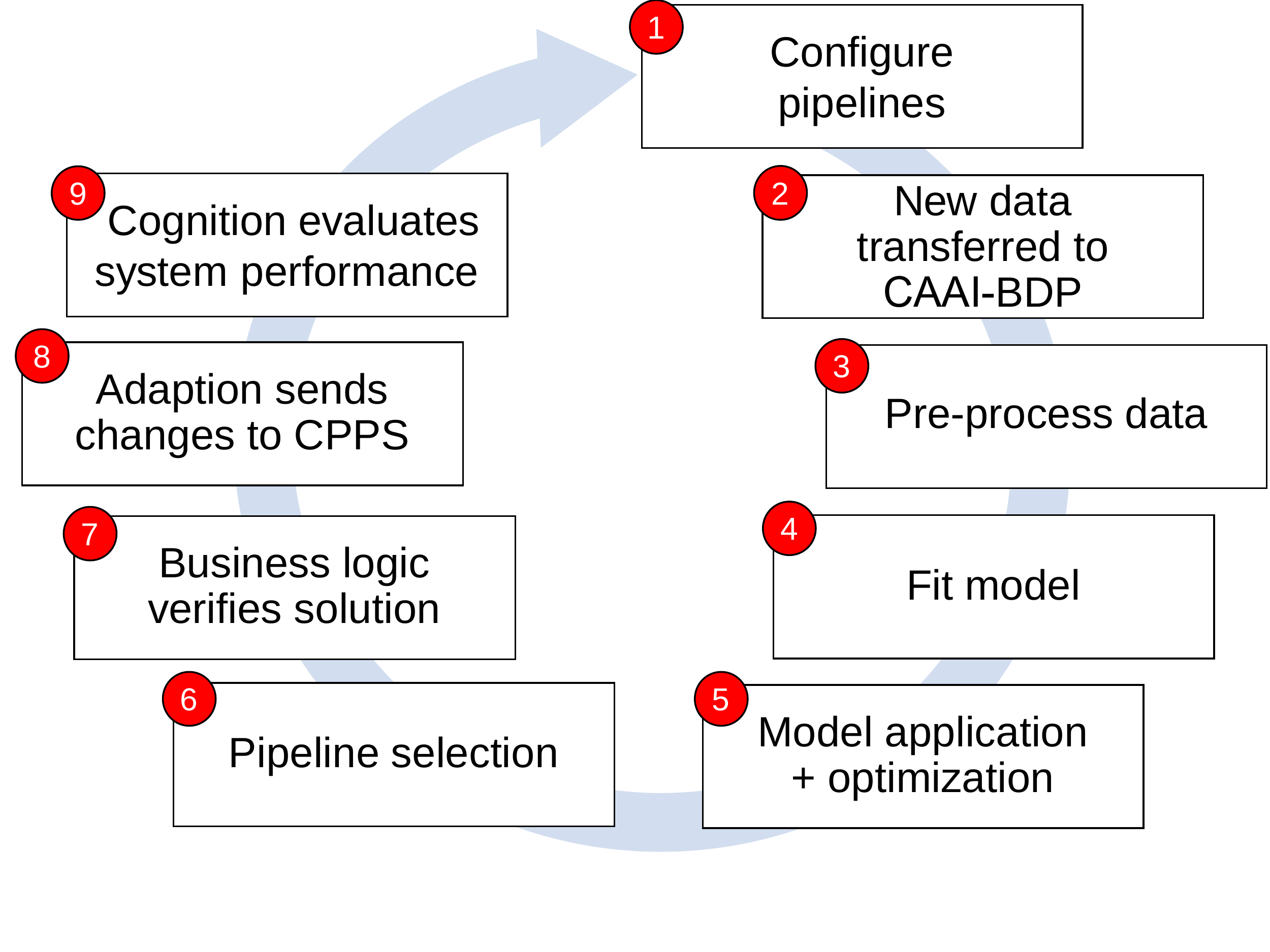}
	\caption{The workflow represent nine steps that are continuously performed to adapt the pipelines and increase their performance over time.}
	\label{fig:workflow}
\end{figure}

\begin{figure*}
	\centering
	\includegraphics[width=0.6\textwidth]{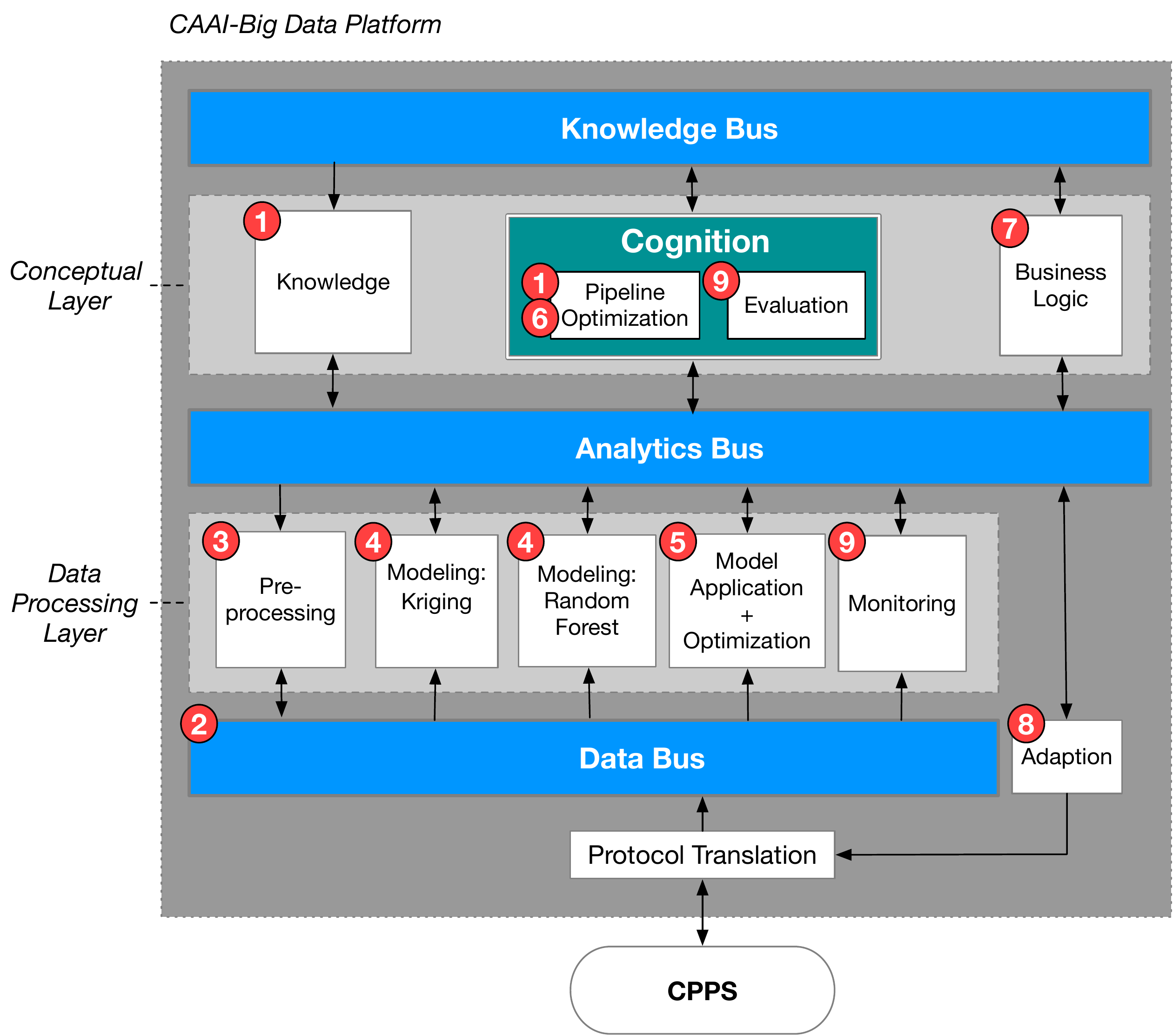}
	\caption{The resulting \gls{CAAI} architecture for the \gls{VPS} use case. The numbers indicate the sequence of the workflow, where some steps can be computed in parallel, i.e., the two different surrogate models. }
	\label{fig:vpsarchitecture}
\end{figure*}

Our architecture uses to select different algorithms and evaluate their results.
The workflow of the architecture is depicted in Figure~\ref{fig:workflow}. 
Its modularity (M-2) enables a comfortable implementation of a \gls{SMBO} algorithm as the functionality of single modules can be re-used. 
The cognition receives necessary information from the knowledge module and starts the workflow consisting of the following nine steps: 

\begin{enumerate}
	\item The \textit{cognition} initializes candidate pipelines for parallel processing by varying model types and parameters. The knowledge module provides the required information about feasible algorithms and boundary constraints. Suitable models for this use case in our algorithm collection are either Kriging or random forest. 
	\item The \textit{protocol translation} module transfers the data from the \gls{OPCUA} server on the \gls{CPPS} to the data bus on the \gls{CBDP}. 
	\item The \textit{preprocessing} module cleans the raw data. 
	As the data quality is good, the pre-processing in this use case is reduced to data normalization. 
	Figure~\ref{fig:messaging} shows the \textit{preprocessing} module publishing messages to the data bus for further (parallel) processing. 
	\begin{figure}
		\centering
		\includegraphics[width=\columnwidth]{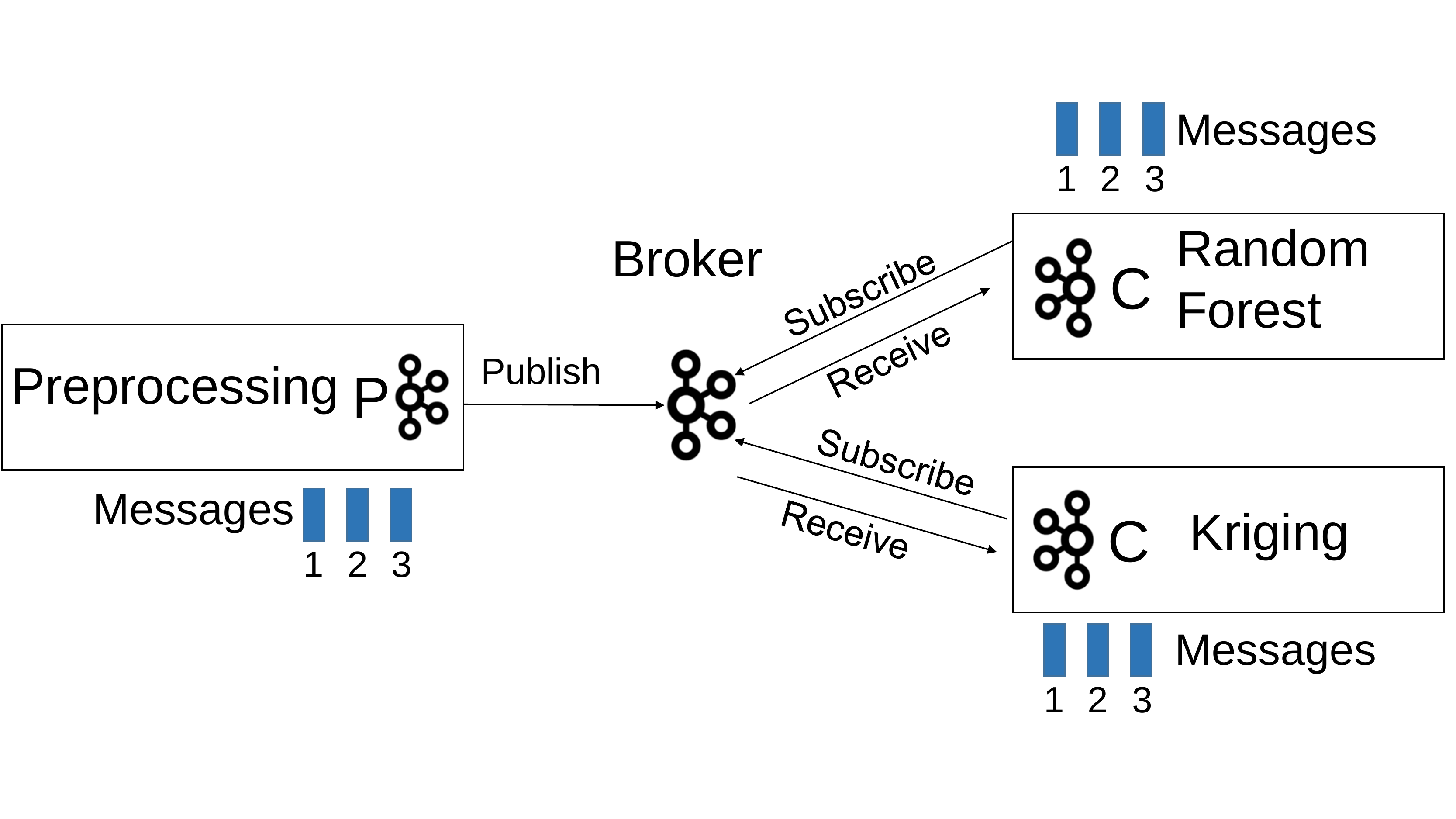}
		\caption{Parallel processing of messages through different algorithms. The \textit{cognition} module instantiated two pipelines with candidate algorithms and assigned them into different consumer groups.
			As two consumer groups are subscribed to this topic, both groups receive all new messages and various algorithms can be trained independently.
			More instances can be added to the same consumer group, which results in a distributed processing of incoming messages.}
		\label{fig:messaging}
	\end{figure}
	\item The \textit{Kriging} and \textit{random forest} model learning components fit or update their parameters and send the results to the analytics bus.
	\item The module \textit{model application + optimization} implements the sequential step of the \gls{SMBO} algorithm: it searches the previously fitted model until an optimal solution is found or the maximum number of iterations is reached. 
	The module transfers the result to the analytics bus.
	\item The \textit{cognition} decides using the model accuracy and predicted optimum, which pipeline will be chosen.
	\item The \textit{business logic} module verifies if the solution violates any of the constraints, e.g., too much corn in the reactor, and communicates the appropriate adaption back to the analytics bus.
	\item The \textit{adaption} module translates the adjustments for the specific \gls{CPPS} and sends the instructions from the \gls{CBDP} to the \gls{CPPS}.
	\item The \textit{cognition} module analyses the system performance, as achieved with the resulting pipeline configuration from step 6. In the following steps, the impact of changes is verified through information provided by the \textit{monitoring} module.
\end{enumerate}
The resulting implementation of the \gls{CAAI} for the given use case, including all applied modules and the described workflow, is illustrated in Figure~\ref{fig:vpsarchitecture}.

\subsection{Results}\label{sec:results}
Data from the real-world \gls{VPS} was acquired to evaluate the modeling and optimization.
This data consists of 36 production cycles with 12 different settings for the runtime of the conveyor.
Based on this data, we trained a model that reflects the real behavior of the \gls{VPS} and utilize it for further experiments.
The three different objectives, i.e., the energy consumption, the processing time, and the amount of corn needed (see Section~\ref{sec:usecase} for more details), were aggregated by taking the sum of the single objectives multiplied with equal weights of $1/3$. 
\begin{figure}
	\includegraphics[width=\columnwidth]{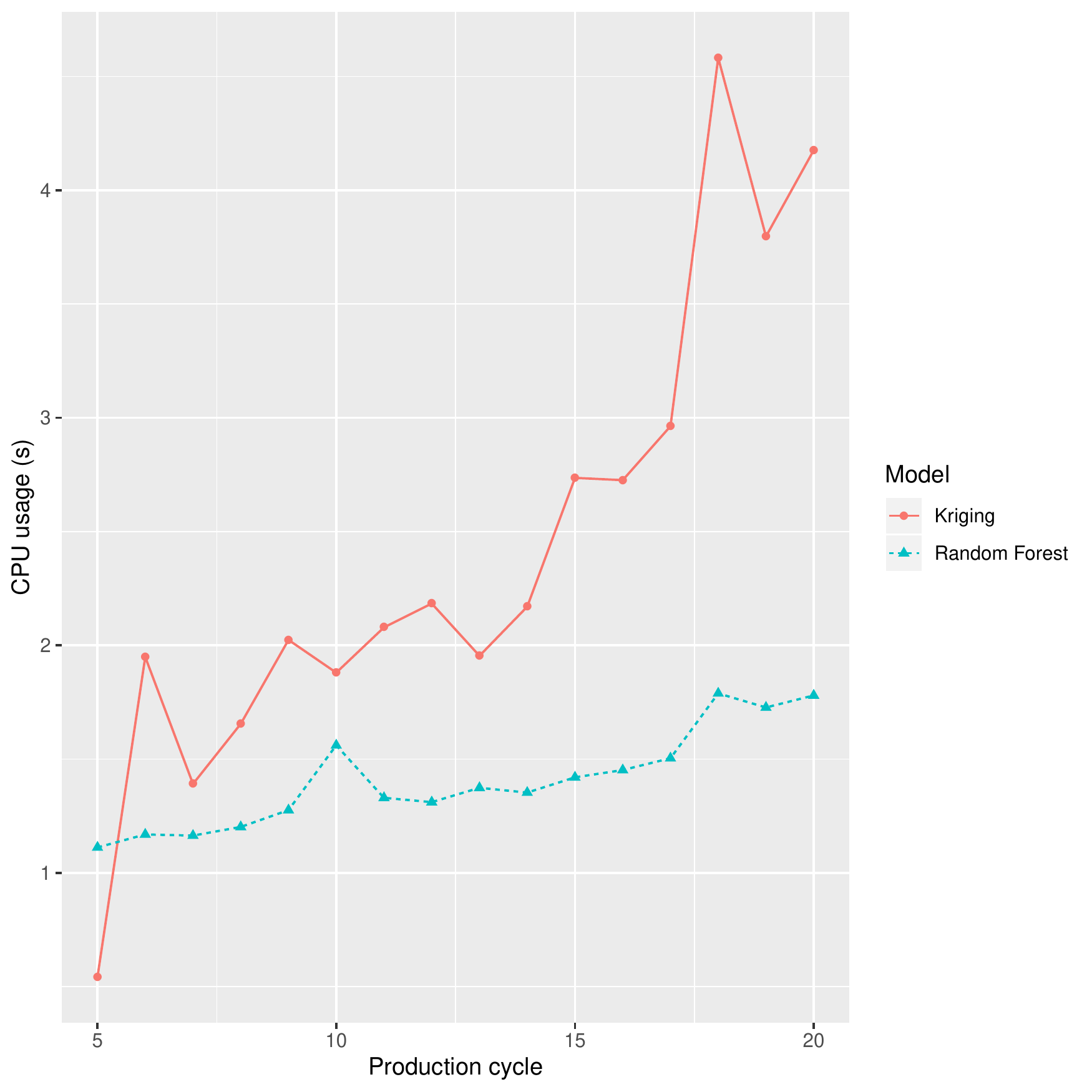}
	\caption{This plot shows the consumed CPU time in seconds over the production cycle, which is equal to the number of data points used. The results are aggregated over 10 repetitions using median values.}
	\label{fig:cpu}
\end{figure}

A MacBook Pro with a quad-core Intel Core i7 CPU at 3.1 GHz and 16 GB DDR3 RAM computed the results described in this Section. 
R version 3.4.3 was the software platform~\cite{R17a} to evaluate the algorithms.
Kriging and random forest employ the software packages SPOT (2.0.5)~\cite{Bart17parxiv} and caret (6.0-84)~\cite{Kuhn08a}.
In the initial phase, the algorithms used five equidistant data points to build their initial models. 
Consequently, the results in the figures presented in this section start at production cycle number five. 
The aggregated results use the median values of ten repetitions, with 20 production cycles in each repetition.
Figure~\ref{fig:cpu} plots the CPU consumption in seconds against the \gls{VPS} production cycles. 

For both methods, Kriging and random forest, an increasing trend can be observed. 
However, the computation time of Kriging shows a larger slope, compared to random forest.
Both algorithms behave as expected, stemming from the internal data representation and processing.  
\begin{figure}
	\includegraphics[width=\columnwidth]{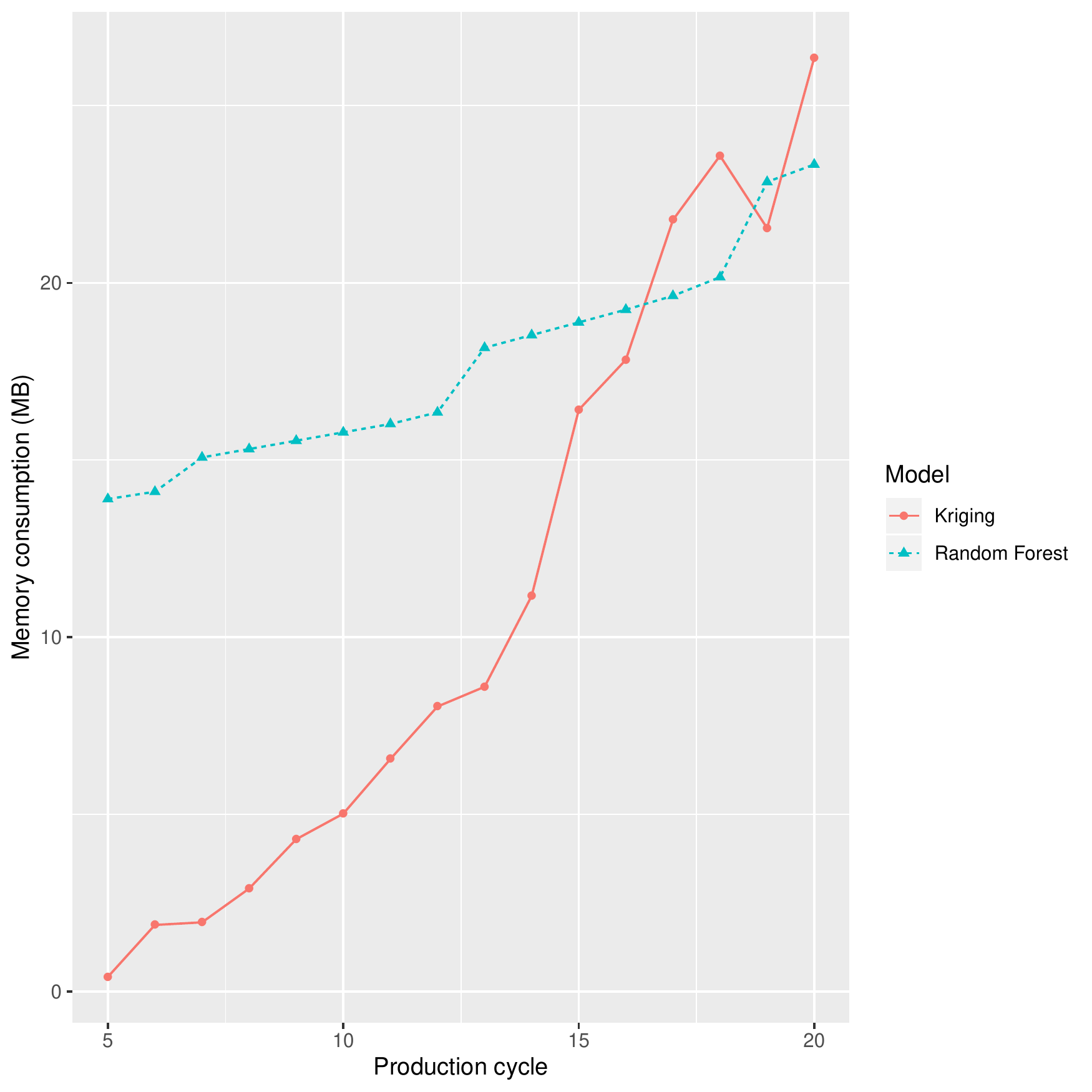}
        \caption{This figure depicts the memory (MB) consumption of the modeling algorithms for different production cycles. The number of the production cycle is equal to the number of data points used in this cycle. 
	The results are aggregated over 10 repetitions using median values.}
	\label{fig:memory}
\end{figure}
The same holds for memory consumption, as depicted in Figure~\ref{fig:memory}.
At the early stage, the random forest algorithm uses more RAM than Kriging. 
After about 15 iterations the situation changes as Kriging started to acquire more memory than random forest. 
While the required memory grows further for both algorithms, Kriging also shows the steeper slope.
\begin{figure}
        \centering
        		\includegraphics[width=\columnwidth]{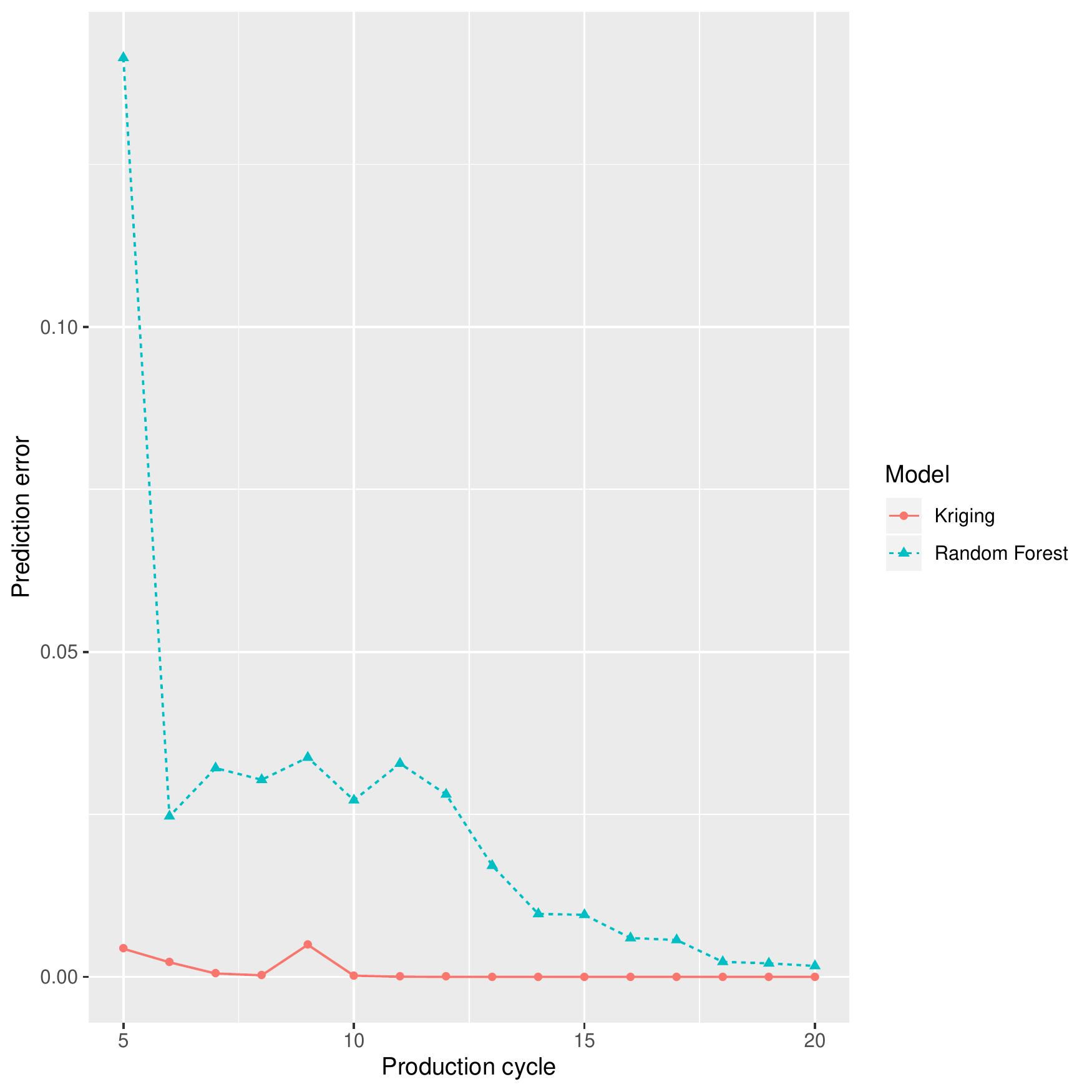}
		\caption{Prediction error plotted against the number of production cycles. 
		The prediction error is the absolute difference between the model prediction of the best found point and the true objective function value.}
		\label{fig:predError}
\end{figure}

Comparing the prediction accuracy of the models at their best-predicted points with the real objective function value, as shown in Figure~\ref{fig:predError}, Kriging shows a nearly constant accurate performance, while random forest shows a larger variance and starts to get comparably accurate predictions in the last production cycles.  

\begin{figure}
        \centering
		\includegraphics[width=\columnwidth]{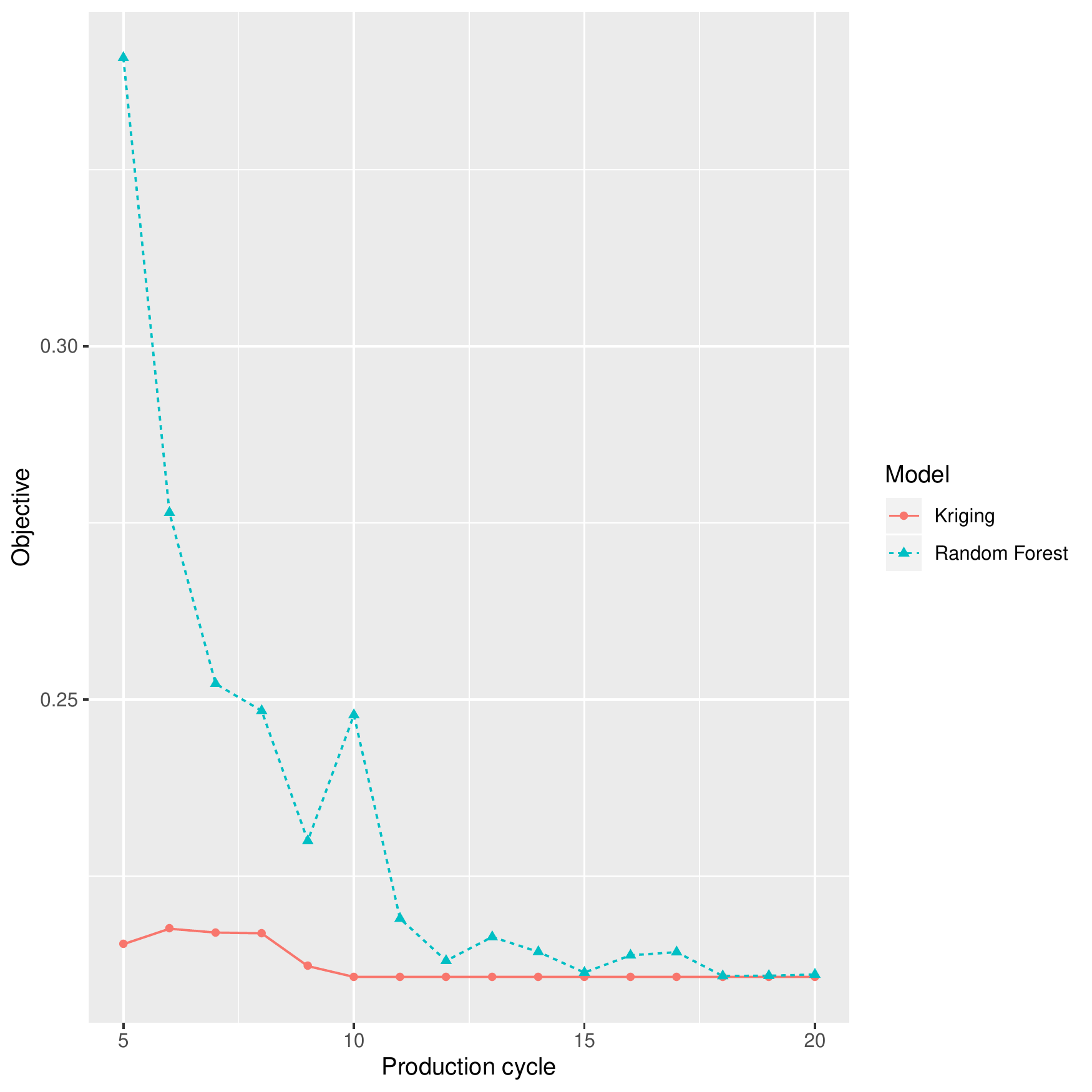}
		\caption{Objective function value plotted against the number of production cycles.}
		\label{fig:y}
\end{figure}

The reached values of the objective function are depicted in Figure~\ref{fig:y}. 
It shows that in the beginning, Kriging outperforms random forest, while later after about 12 cycles, random forests perform comparably to Kriging.
Results from our study indicate the following, valuable findings:
\begin{enumerate} [(i)]
	\item It is worth using more than one algorithm: taking only the best performing (i.e., Kriging) can lead to future problems due to limited computation resources and time.
	\item Random forest needs more data to improve compared to Kriging and starts to be a good competitor after about 15 cycles.
	\item The prediction error is a useful measure to detect performance drifts and switch to other pipelines if needed.
	\item Altogether, it is beneficial to switch algorithms after a certain number of production cycles, when regarding all performance metrics together.
	\item A forgetting mechanism is necessary to implement \gls{SMBO} efficiently for long term usage in \gls{CPPS} scenarios due to physically limited computation resources.  
	This mechanism could be a fixed or adaptive size of the model, e.g., the sliding window approach applied in time series computation, and a method to choose which data to remove from the model. 
\end{enumerate}
Additionally, the processed material can change its behavior over time. 
Storing the corn over an extended period eventually requires more corn to fill the cups with the desired amount.
Therefore it is crucial to adapt the system during runtime by changing the computation pipelines, recomputing models, and adjusting model sizes.

\section{Conclusion and Outlook}\label{sec:conclusion}
\begin{figure*}
	\centering
	\includegraphics[width=0.6\textwidth]{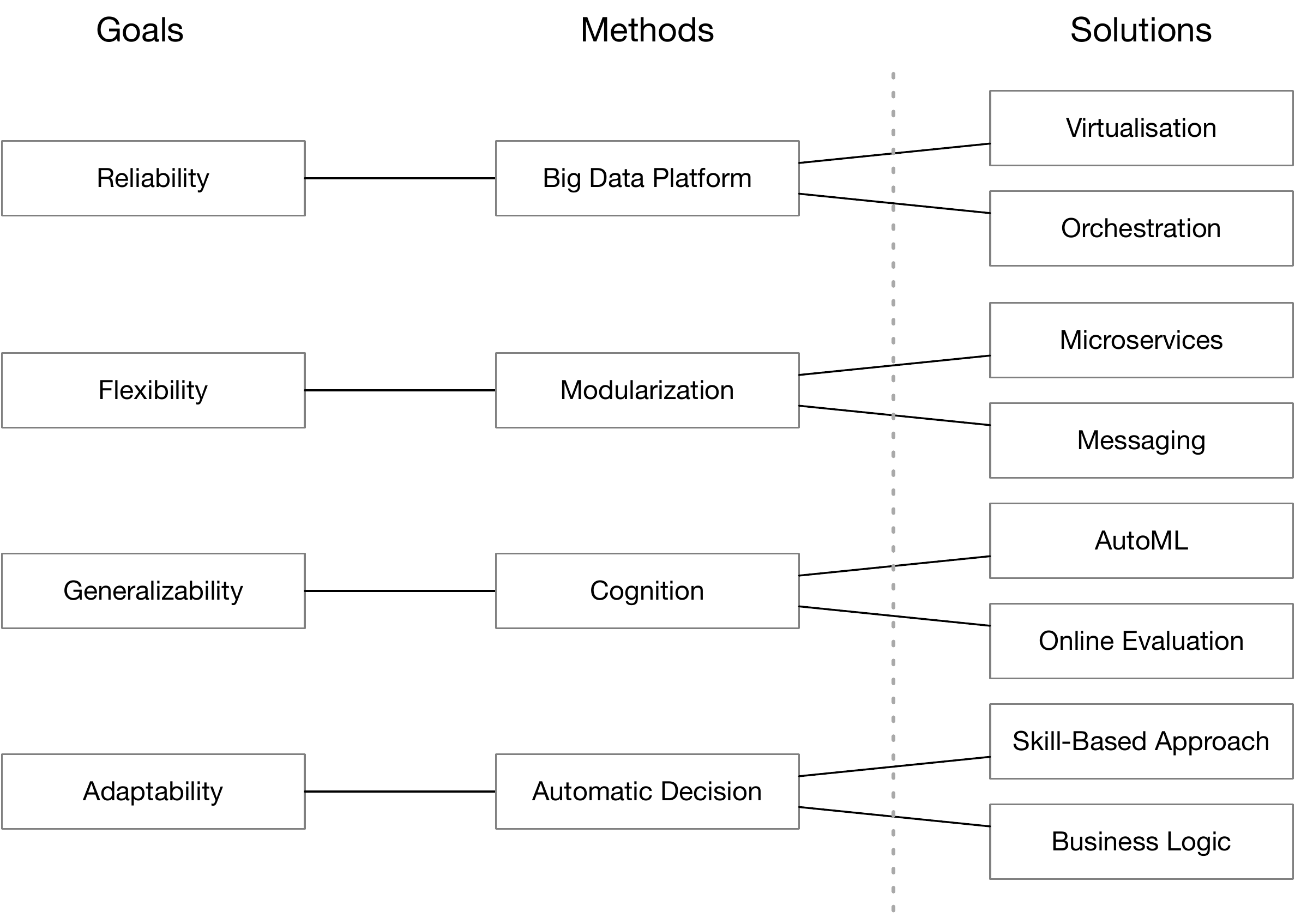}
	\caption{Overview of goals, methods and solutions. The goals and methods on the left side represent the parts with a larger reference character of the architecture, while the solutions on the right side rather represent the current state-of-the-art in the respective disciplines.}
	\label{fig:goaloverview}
\end{figure*}

%
%
In this paper, we defined defined goals (G1-G4) to be reached by a cognitive architecture to improve or maintain the efficiency of a CPPS. 
Each goal is addressed by a particular method (M1-M4), which can be implemented by several technologies, e.g., \mbox(T1)-\mbox{(T4)} or solutions. 
Figure~\ref{fig:goaloverview} details the coherence of these goals, methods, and solutions which results in our cognitive architecture \gls{CAAI}, which was presented, and further evaluated on a real-world problem in this work. 
Different manifestations of this architecture are possible, with one implementation being examplified for the VPS use case evaluation.
The key feature is the cognition module that configures, instantiates and evolves process pipelines over time to solve the problem, i.e., to reach the addressed goal. 

%
%
Besides the main contribution of this paper, the proposed \gls{CAAI}, the major results (R) can be summarized as follows:
\begin{enumerate}[(R-1)]
	\item The central aim, to efficiently optimize the performance of a CPPS, can not be reached by addressing single goals or implementing single methods individually (see Section~\ref{sec:introduction}).
	The main implication is that the individual goals support each other to a high degree.
	For example, a flexible and modular system is required by the cognition component to allow generalization of the system and to be applicable to different use cases.
	\item 
	The defined goals and methods represent the reference character of the proposed architecture and will maintain their validity over a long time.
	However, the solutions and chosen technologies or concepts for implementation may be subject to change in the future. 
	For example, Docker is a valid choice to fulfill the requirements of CPPS scenarios, to process data in near real-time (see Section~\ref{sec:implementation}). 
	As technologies evolve, this solution may be replaced by a more modern one in the future but can be regarded as a state-of-the-art virtualization method nowadays. 
	\item The performance of an algorithm for a given use case and data may change over time, in both directions (worsening and improving) (see Section~\ref{sec:results}). 
	Therefore, an offline benchmark and selection, which chooses one particular algorithm as the best performing might be misleading. 
	Moreover, the processing time of an particular algorithm can significantly increase due to growing data volume, e.g., through additional sensors in the CPPS.
	Hence, changing the processing pipeline to another algorithm can be beneficial to reduce computational resources costs.
	To realize such an online selection and establish an efficient system, the cognitive module needs sufficient degrees of freedom and thus a somewhat broad portfolio of algorithms. 
	\item Overall, the \gls{CAAI} is specific enough to support concrete implementation in form of the \gls{CBDP}. 
	The resulting system was able to optimize a simple I4.0 use case by configuring, instantiating, and evaluating several processing pipelines.   
\end{enumerate}
From our point of view, these outcomes altogether cover three important disciplines in the field of AI-applications and the related methods: 
\begin{enumerate}[(i)]
	\item \textbf{Infrastructure:} Must be reliable and flexible to fulfill response time requirements in industry scenarios \mbox{(M-1 and M-2)}. 
	\item \textbf{Learning:} Suitable algorithm portfolio and topology needed to address several declarative goals and resource limitations \mbox{(M-3)}. 
	\item \textbf{Data curation:} Combines data pre-processing and domain knowledge to create metadata to support further algorithmic processing (\mbox{M-3} and \mbox{M-4}). 
\end{enumerate}

%
%
The work on this paper reveals some questions, leading to logical next steps and follow-up research tasks: 
The effort to solve a different use case by implementing and adjusting the \gls{CAAI}, which should be minimized, has to be analyzed. 
This would directly take up results \mbox{(R-2)}, regarding the reference character of the \gls{CAAI}, and \mbox{(R-3)}, regarding the algorithm portfolio. 
Additionally, e.g., configuration for the orchestration, time requirements, or data pre-processing, may change as well. 

The implementation of pipeline creation and evaluation is adequate for the presented, in terms of the optimization problem rather straightforward, use case. 
However, further efforts are required to build a truly intelligent system that can solve harder use cases through learning over time and re-calibration in an online manner \mbox{(R-3)}. 
Automatically adapting the CPPS was not yet considered for the regarded use case, as we focus in particular on the \gls{CBDP} and the cognitive module.
Therefore, our future work will include a representation of the necessary knowledge to safely change the configuration and operation of the CPPS while trying to increase its efficiency.

To give an idea of a possible long-term perspective, the degree of cognition addressed by our architecture can be the topic of future research.  
We restricted the scope of the cognition to typical \gls{I4.0} use cases, see Definition~\ref{def:cognition}. 
An extension of the cognitive capabilities to detect restrictions (e.g., introduced by the provided algorithms or the limited available resources) independently and to learn how to deal with them autonomously would be the next step towards a self-aware system. 
\begin{acknowledgements}
The work was supported by the German Federal Ministry of Education and Research (BMBF) under the project "KOARCH" (funding code: 13FH007IA6 and 13FH007IB6).
\end{acknowledgements}

%
\section*{Conflict of interest}
The authors declare that they have no conflict of interest.

\bibliographystyle{spmpsci}      
\bibliography{Literature}

\begin{thebibliography}{10}
\providecommand{\url}[1]{{#1}}
\providecommand{\urlprefix}{URL }
\expandafter\ifx\csname urlstyle\endcsname\relax
  \providecommand{\doi}[1]{DOI~\discretionary{}{}{}#1}\else
  \providecommand{\doi}{DOI~\discretionary{}{}{}\begingroup
  \urlstyle{rm}\Url}\fi

\bibitem{VDI:2015}
Adolphs, P., et~al.: {Reference Architecture Model Industrie 4.0 (RAMI4.0)}.
\newblock VDI /VDE Statusreport p.~32 (2015).
\newblock
  \urlprefix\url{https://www.zvei.org/fileadmin/user_upload/Themen/Industrie_4.0/Das_Referenzarchitekturmodell_RAMI_4.0_und_die_Industrie_4.0-Komponente/pdf/5305_Publikation_GMA_Status_Report_ZVEI_Reference_Architecture_Model.pdf}.
\newblock Accessed: 2020-02-18

\bibitem{Anderson:1996}
Anderson, J.R.: {A Simple Theory of Complex Cognition}.
\newblock American Psychologist  (1996).
\newblock \doi{10.1037/0003-066X.51.4.355}

\bibitem{Bart17parxiv}
Bartz-Beielstein, T., Gentile, L., Zaefferer, M.: In a nutshell: Sequential
  parameter optimization.
\newblock ArXiv e-prints \textbf{1712.04076} (2017)

\bibitem{Bart16n}
Bartz-Beielstein, T., Zaefferer, M.: Model-based methods for continuous and
  discrete global optimization.
\newblock Applied Soft Computing \textbf{55}, 154 -- 167 (2017).
\newblock \doi{10.1016/j.asoc.2017.01.039}

\bibitem{VBS:2017}
Bauer, K., Diemer, J., Hilger, C., L\"{o}wen, U., Michels, J.S.: Benefits of
  application scenario value-based service.
\newblock Tech. rep., Federal Ministry for Economic Affairs and Energy (BMWi)
  (2017)

\bibitem{Brei01a}
Breiman, L.: {Random Forests}.
\newblock Machine Learning \textbf{45}(1), 5--32 (2001)

\bibitem{BFS:2019}
Bunte, A., Fischbach, A., Strohschein, J., Bartz-Beielstein, T.,
  Faeskorn-Woyke, H., Niggemann, O.: Evaluation of cognitive architectures for
  cyber-physical production systems.
\newblock In: 24nd IEEE International Conference on Emerging Technologies and
  Factory Automation (ETFA), pp. 729--736. Zaragoza, Spain (2019).
\newblock \doi{10.1109/ETFA.2019.8869038}

\bibitem{Bunte:2019a}
Bunte, A., Wunderlich, P., Moriz, N., Li, P., Mankowski, A., Rogalla, A.,
  Niggemann, O.: Why symbolic ai is a key technology for self-adaption in the
  context of cpps.
\newblock In: 24nd IEEE International Conference on Emerging Technologies and
  Factory Automation (ETFA). Zaragoza, Spain (2019)

\bibitem{Conf20a}
Confluent: Schema management (2020).
\newblock
  \urlprefix\url{https://docs.confluent.io/current/schema-registry/index.html}.
\newblock Accessed: 2020-02-18

\bibitem{Drath:2014}
{Drath}, R., {Horch}, A.: Industrie 4.0: Hit or hype? [industry forum].
\newblock IEEE Industrial Electronics Magazine \textbf{8}(2), 56--58 (2014).
\newblock \doi{10.1109/MIE.2014.2312079}

\bibitem{FOF:2013}
{European Factories of the Future Research Association (EFFRA)}: Factories of
  the Future: Multi-annual Roadmap for the Contractual PPP Under Horizon 2020.
\newblock EDC collection. Publications Office of the European Union (2013).
\newblock \urlprefix\url{https://books.google.de/books?id=ZC0wngEACAAJ}.
\newblock Accessed: 2020-02-18

\bibitem{Feur15a}
Feurer, M., Klein, A., Eggensperger, K., Springenberg, J., Blum, M., Hutter,
  F.: Efficient and robust automated machine learning.
\newblock In: C.~Cortes, N.D. Lawrence, D.D. Lee, M.~Sugiyama, R.~Garnett
  (eds.) Advances in Neural Information Processing Systems 28, pp. 2962--2970.
  Curran Associates, Inc. (2015)

\bibitem{Fusi18a}
Fusi, N., Sheth, R., Elibol, M.: Probabilistic matrix factorization for
  automated machine learning.
\newblock In: Advances in Neural Information Processing Systems, pp. 3348--3357
  (2018)

\bibitem{Gokalp2016}
G{\"{o}}kalp, M.O., Kayabay, K., Akyol, M.A., Eren, P.E., Ko{\c{c}}yigit, A.:
  {Big Data for Industry 4.0 : A Conceptual Framework}.
\newblock 2016 International Conference on Computational Science and
  Computational Intelligence (December) (2016).
\newblock \doi{10.1109/CSCI.2016.87}

\bibitem{High17a}
Hightower, K., Burns, B., Beda, J.: Kubernetes: Up and Running Dive into the
  Future of Infrastructure, 1st edn.
\newblock O'Reilly Media, Inc. (2017)

\bibitem{ibmmq}
{IBM Knowledge Center}: Main features and benefits of message queuing (2019).
\newblock
  \urlprefix\url{https://www.ibm.com/support/knowledgecenter/SSFKSJ\_9.1.0/com.ibm.mq.pro.doc/q002630\_.htm}.
\newblock Accessed: 2020-02-18

\bibitem{Jin05a}
Jin, Y.: A comprehensive survey of fitness approximation in evolutionary
  computation.
\newblock Soft computing \textbf{9}(1), 3--12 (2005)

\bibitem{Jung17a}
Jung, C., Zaefferer, M., Bartz-Beielstein, T., Rudolph, G.: Metamodel-based
  optimization of hot rolling processes in the metal industry.
\newblock The International Journal of Advanced Manufacturing Technology
  \textbf{90}(1), 421--435 (2017).
\newblock \doi{10.1007/s00170-016-9386-6}.
\newblock \urlprefix\url{https://doi.org/10.1007/s00170-016-9386-6}

\bibitem{Kagermann:2013}
Kagermann, H., Wahlster, W., Helbig, J.: {Securing the future of German
  manufacturing industry Recommendations for implementing the strategic
  initiative INDUSTRIE 4.0}.
\newblock acatech -- National Academy of Science and Engineering, Berlin (2013)

\bibitem{Krige1951}
Krige, D.: A statistical approach to some basic mine valuation problems on the
  witwatersrand.
\newblock Journal of the Chemical, Metallurgical and Mining Society of South
  Africa \textbf{52}(6), 119--139 (1951)

\bibitem{Kuhn08a}
Kuhn, M.: Building predictive models in r using the caret package.
\newblock Journal of Statistical Software, Articles \textbf{28}(5), 1--26
  (2008).
\newblock \doi{10.18637/jss.v028.i05}.
\newblock \urlprefix\url{https://www.jstatsoft.org/v028/i05}

\bibitem{Laird:1987}
Laird, J.E., Newell, A., Rosenbloom, P.S.: Soar: An architecture for general
  intelligence.
\newblock Artif. Intell. \textbf{33}(1), 1--64 (1987).
\newblock \doi{10.1016/0004-3702(87)90050-6}.
\newblock \urlprefix\url{http://dx.doi.org/10.1016/0004-3702(87)90050-6}

\bibitem{Lee:2017}
Lee, J., Jin, C., Bagheri, B.: Cyber physical systems for predictive production
  systems.
\newblock Production Engineering \textbf{11}(2), 155--165 (2017).
\newblock \doi{10.1007/s11740-017-0729-4}.
\newblock \urlprefix\url{https://doi.org/10.1007/s11740-017-0729-4}

\bibitem{Li:2019}
Li, D., Fast-Berglund, {\AA}., Paulin, D.: Current and future industry 4.0
  capabilities for information and knowledge sharing.
\newblock The International Journal of Advanced Manufacturing Technology
  \textbf{105}(9), 3951--3963 (2019).
\newblock \doi{10.1007/s00170-019-03942-5}.
\newblock \urlprefix\url{https://doi.org/10.1007/s00170-019-03942-5}

\bibitem{Lin2017}
Lin, S.W., et~al.: {The Industrial Internet of Things Volume G1: Reference
  Architecture v1.80}.
\newblock Tech. rep., Industrial Internet Consortium (2017)

\bibitem{Malakuti:2018}
{Malakuti}, S., {Bock}, J., {Weser}, M., {Venet}, P., {Zimmermann}, P.,
  {Wiegand}, M., {Grothoff}, J., {Wagner}, C., {Bayha}, A.: Challenges in
  skill-based engineering of industrial automation systems*.
\newblock In: 2018 IEEE 23rd International Conference on Emerging Technologies
  and Factory Automation (ETFA), vol.~1, pp. 67--74 (2018).
\newblock \doi{10.1109/ETFA.2018.8502635}

\bibitem{Marler2010}
Marler, R., Arora, J.: The weighted sum method for multi-objective
  optimization: New insights.
\newblock Structural and Multidisciplinary Optimization \textbf{41}, 853--862
  (2010).
\newblock \doi{10.1007/s00158-009-0460-7}

\bibitem{Nark17a}
Narkhede, N., Shapira, G., Palino, T.: Kafka: The Definitive Guide Real-Time
  Data and Stream Processing at Scale, 1st edn.
\newblock O'Reilly Media, Inc. (2017)

\bibitem{Negu15a}
Negus, C.: Docker Containers, 2nd edn.
\newblock Addison-Wesley Professional (2015)

\bibitem{Neis14a}
Neisser, U.: Cognitive psychology: Classic edition.
\newblock Psychology Press (2014)

\bibitem{Olso16a}
Olson, R.S., Urbanowicz, R.J., Andrews, P.C., Lavender, N.A., Kidd, L.C.,
  Moore, J.H.: Applications of Evolutionary Computation: 19th European
  Conference, EvoApplications 2016, Porto, Portugal, March 30 -- April 1, 2016,
  Proceedings, Part I, chap. Automating Biomedical Data Science Through
  Tree-Based Pipeline Optimization, pp. 123--137.
\newblock Springer International Publishing (2016).
\newblock \doi{10.1007/978-3-319-31204-0_9}.
\newblock \urlprefix\url{http://dx.doi.org/10.1007/978-3-319-31204-0_9}

\bibitem{Bmwi:2019}
{Plattform Industrie 4.0}: {Technology Scenario ‘Artificial Intelligence in
  Industrie 4.0’}  (2019).
\newblock
  \urlprefix\url{https://www.plattform-i40.de/PI40/Redaktion/EN/Downloads/Publikation/AI-in-Industrie4.0.pdf}.
\newblock Accessed: 2020-02-18

\bibitem{R17a}
{R Core Team}: R: A Language and Environment for Statistical Computing.
\newblock R Foundation for Statistical Computing, Vienna, Austria (2017).
\newblock \urlprefix\url{https://www.R-project.org/}.
\newblock Accessed: 2020-02-18

\bibitem{Santos2017}
Santos, M.Y., S{\'{a}}, J.O., Costa, C., Galvao, J., Andrade, C., Martinho, B.,
  Lima, F.V., Costa, E.: {A Big Data Analytics Architecture for Industry 4.0}.
\newblock Advances in Intelligent Systems and Computing  (2017).
\newblock \doi{10.1007/978-3-319-56538-5_19}

\bibitem{Schroeder:2016}
Schröder, C.: The Challenges of Industry 4.0 for Small and Medium-sized
  Enterprises (2016).
\newblock \urlprefix\url{http://library.fes.de/pdf-files/wiso/12683.pdf}.
\newblock Accessed: 2020-02-18

\bibitem{Thom18a}
Thomas, J., Coors, S., Bischl, B.: Automatic gradient boosting.
\newblock ArXiv e-prints \textbf{1807.03873} (2018)

\bibitem{Thor13a}
Thornton, C., Hutter, F., Hoos, H.H., Leyton-Brown, K.: Auto-weka: Combined
  selection and hyperparameter optimization of classification algorithms.
\newblock In: Proceedings of the 19th ACM SIGKDD International Conference on
  Knowledge Discovery and Data Mining, KDD '13, pp. 847--855. ACM, New York,
  NY, USA (2013).
\newblock \doi{10.1145/2487575.2487629}.
\newblock \urlprefix\url{http://doi.acm.org/10.1145/2487575.2487629}

\bibitem{Watt15a}
Watts, K.: Microservices Architecture: Deep Exploration Of Microservices.
\newblock CreateSpace Independent Publishing Platform, North Charleston, SC,
  USA (2015)

\bibitem{Weskamp:2019}
Weskamp, J.N., Pethig, F., Al-gumaei, K., Poudel, B.K.: {Offene
  Big-Data-Plattform} (March) (2019)

\bibitem{Xing18a}
Xing, B., Xiao, Y., Qin, Q.H., Cui, H.: Quality assessment of resistance spot
  welding process based on dynamic resistance signal and random forest based.
\newblock The International Journal of Advanced Manufacturing Technology
  \textbf{94}(1), 327--339 (2018).
\newblock \doi{10.1007/s00170-017-0889-6}.
\newblock \urlprefix\url{https://doi.org/10.1007/s00170-017-0889-6}

\bibitem{Zimmermann:2019}
{Zimmermann}, P., {Axmann}, E., {Brandenbourger}, B., {Dorofeev}, K.,
  {Mankowski}, A., {Zanini}, P.: Skill-based engineering and control on
  field-device-level with opc ua.
\newblock In: 2019 24th IEEE International Conference on Emerging Technologies
  and Factory Automation (ETFA), pp. 1101--1108 (2019).
\newblock \doi{10.1109/ETFA.2019.8869473}

\end{thebibliography}
\end{document}